\documentclass[doublecol]{epl2}
\usepackage{amssymb,amsmath,graphicx,setspace,color,rotating,subfigure,url}
\usepackage{extarrows}
\usepackage{CJK}
\title{Quantifying immediate price impact of trades based on the $k$-shell decomposition of stock trading networks}
\shorttitle{Quantifying immediate price impact of traders} 

\author{Wen-Jie Xie\inst{1,2,3} \and Ming-Xia Li\inst{2,3,4} \and Hai-Chuan Xu\inst{1,2,3} \and Wei Chen\inst{5} \and Wei-Xing Zhou\inst{1,3,6}\footnote{e-mail: wxzhou@ecust.edu.cn} \and H. Eugene Stanley\inst{7}}
\shortauthor{W.-J. Xie \etal}

\institute{

  \inst{1} Department of Finance, East China University of Science and Technology, Shanghai 200237, China\\
  \inst{2} Postdoctoral Research Station, East China University of Science and Technology, Shanghai 200237, China\\
  \inst{3} Research Center for Econophysics, East China University of Science and Technology, Shanghai 200237, China\\
  \inst{4} School of Sports Science and Engineering, East China University of Science and Technology, Shanghai 200237, China\\
  \inst{5} Shenzhen Stock Exchange, 5045 Shennan East Road, Shenzhen 518010, China\\
  \inst{6} Department of Mathematics, East China University of Science and Technology, Shanghai 200237, China\\
  \inst{7} Department of Physics and Center for Polymer Studies, Boston  University, Boston, MA 02215, USA\\
}


\abstract{
Traders in a stock market exchange stock shares and form a stock trading network. Trades at different positions of the stock trading network may contain different information. We construct stock trading networks based on the limit order book data and classify traders into $k$ classes using the $k$-shell decomposition method. We investigate the influences of trading behaviors on the price impact by comparing a closed national market (A-shares) with an international market (B-shares), individuals and institutions, partially filled and filled trades, buyer-initiated and seller-initiated trades, and trades at different positions of a trading network. Institutional traders professionally use some trading strategies to reduce the price impact and individuals at the same positions in the trading network have a higher price impact than institutions. We also find that trades in the core have higher price impacts than those in the peripheral shell.
}

\begin{document}
\maketitle



\section{Introduction}

The availability of large-scale data on economic and financial activities provides great challenges and new opportunities for us to gain a deeper understanding of the dynamics of complex economic and financial systems \cite{Einav-Levin-2014-Science,Bouchaud-2008-Nature}, in which the structure and evolutionary dynamics of complex economic and financial networks play an essential role \cite{Schweitzer-Fagiolo-Sornette-VegaRedondo-Vespignani-White-2009-Science,Pukthuanthong-Roll-2009-JFE,Schiavo-Reyes-Fagiolo-2010-QF,Battiston-Farmer-Flache-Garlaschelli-Haldane-Heesterbeek-Hommes-Jaeger-May-Scheffer-2016-Science}.
In financial and economic networks, the nodes represent financial or economic agents, such as economies, companies, financial institutions, traders, et al., while the links represent interactions between two nodes, such as investment, trade, lending, economic cooperation, and so on \cite{Boss-Elsinger-Summer-Thurner-2004-QF,Garlaschelli-Loffredo-2004-PRL,Hochberg-Ljungqvist-Lu-2007-JF,Kogut-Urso-Wakler-2007-MS,Fagiolo-Reyes-Schiavo-2009-PRE}.


The buy-sell interactions among traders in economic systems can be described by trading networks, in which the nodes represent the traders and the edges stand for the trading relationships. The main statistical properties of several trading networks have been investigated, such as the Austrian money flow trading network \cite{Kyriakopoulos-Thurner-Puhr-Schmitz-2009-EPJB}, the trading network in a web-based experimental prediction market \cite{Tseng-Li-Wang-2010-EPJB, Tseng-Lin-Lin-Wang-Li-2010-PA}, the daily trading networks in the Shenzhen stock market \cite{Jiang-Zhou-2010-PA}, and the trading networks in the Shanghai Futures Market \cite{Wang-Zhou-Guan-2011-PA}. Usually, these trading networks are scale-free with power-law degree distributions and disassortative.

The statistical properties of trading networks can be utilized to track and detect abnormal trades implemented by price manipulators in financial markets. Kyriakopoulos et al. performed random matrix analysis to identify accounts with financial misconduct \cite{Kyriakopoulos-Thurner-Puhr-Schmitz-2009-EPJB}, Tumminello et al. identified trader clusters with a very high degree of synchronization in trading which implies to some extent the presence of price manipulation \cite{Tumminello-Lillo-Piilo-Mantegna-2012-NJP}, Sun et al. found significant differences in the topological properties between manipulated and non-manipulated stocks \cite{Sun-Cheng-Shen-Wang-2011-PA,Sun-Shen-Cheng-Wang-2012-PLoS1}, and Jiang et al. studied the behavior and implications of abnormal trading motifs (self-loop, two-node loop, and two-node multiple arcs) \cite{Jiang-Xie-Xiong-Zhang-Zhang-Zhou-2013-QFL}. There are also studies on the correlations between structural properties of trading networks and financial variables of markets \cite{Adamic-Brunetti-Harris-Kirilenko-2012-SSRN,Li-Jiang-Xie-Xiong-Zhang-Zhou-2015-PA}. It is also reported that trading networks have predictive power over stock price movements at the daily level \cite{Sun-Shen-Cheng-2014-SR} and traders' returns are correlated with their positions occupied in the trading networks \cite{Cohen-Cole-Kirilenko-Patacchini-2011-SSRN}.

In this Letter, we investigate the immediate price impact of institutional and individual trades at different positions of trading networks. The immediate price impact was extensively studied, which is composed of the positive correlations between trading volume and volatility. These relations are robust at various time  scales \cite{Wood-McInish-Ord-1985-JF,Gallant-Rossi-Tauchen-1992-RFS,Richardson-Sefcik-Thompson-1986-JFE}, even at the transaction level \cite{Chan-Fong-2000-JFE,Lillo-Farmer-Mantegna-2003-Nature,Lim-Coggins-2005-QF,Zhou-2012-QF}. Using the same order book data \cite{Mu-Zhou-Chen-Kertesz-2010-NJP,Zhou-Mu-Kertesz-2012-NJP}, it is found that filled and partially filled limit orders have very different price impacts \cite{Zhou-2012-QF}. The price impact of trades from partially filled orders is constant when the volume is not too large, while that of filled orders shows a power-law behaviour $r=\omega^{\alpha}$ with $\alpha\approx2/3$. Zhou also found that large trade sizes, wide bid-ask spreads, high liquidity at the same side and low liquidity at the opposite side will cause a large price impact \cite{Zhou-2012-NJP}. Using the $k$-shell decomposition method, we extend the analysis of immediate price impact by considering the positions of trades at trading networks.

\section{Construction of stock trading networks from transaction data}

We will continue to investigate immediate price impact of institutional and individual trades by using the order book data in the Shenzhen Stock Exchange (SZSE). The data sets in this paper include the order book data of 32 A-shares and 11 B-shares in the SZSE. The A-share market and The B-share market both are composed of common stocks which are issued by mainland Chinese companies. The A-share market is opened only to domestic investors, and traded in CNY. But the B-share market is traded in Hong Kong dollar (HKD) and was restricted to foreign investors before February 19, 2001, and since then it has been opened to Chinese investors as well. A transaction is triggered by an incoming market order matched with the limit orders waiting on the opposite order book and accomplished by transferring shares from seller to buyer and cashes from buyer to seller. It provides an opportunity to trace the order execution procedure from a complex network perspective.

We use the same approach as presented in Refs.~\cite{Jiang-Zhou-2010-PA,Jiang-Xie-Xiong-Zhang-Zhang-Zhou-2013-QFL,Li-Jiang-Xie-Xiong-Zhang-Zhou-2015-PA} to construct stock trading networks. Firstly we reconstruct the limit order book based on the trading rules and extract the detailed information of each transaction. A node represents a trader who bought or sold the stock. An undirected link is formed between two traders if they had transactions between them. Then we present the trades between pairwise traders into an symmetrical adjacent matrix $A_{n\times n}$ whose element $a_{ij}$ equal to 1 or 0. Its entry $a_{ij}=1$ means that trader $i$ has traded with trader $j$. When a trader places an effective
market order, it is possible that the order is executed by several orders which are submitted by different traders on the limit order book. In this case, the local network structure is a star-like graph.

\begin{figure}[htb]
\centering
\includegraphics[width=8cm]{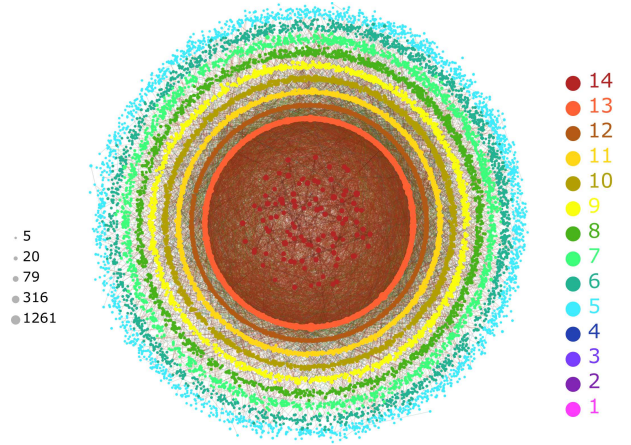}
  \caption{(color online) Topological structure of a stock trading network. Each shell is labelled by a single index $k$ with the nodes colored uniquely. The nodes in the core are red. The size of a node is proportional to its degree. For better visualization, we plot 5\% of the edges and the nodes with $k\geq 5$.}
  \label{Fig:STN_Kshell}
\end{figure}

\section{Trader classification based on $k$-shell decomposition}

Analyzing the undirected unweighted stock trading networks, we classify traders into $k$ shells using the $k$-shell decomposition method. The 1-shell denotes the peripheral shell of the trading network and the $k_{\rm{max}}$-shell denotes the traders in the core of the trading network. Fig.~\ref{Fig:STN_Kshell} shows the $k$-shells of one stock trading network with the LaNet-vi visualization \cite{Beiro-Alvarez-Hamelin-Busch-2008-NJP}, which provides a method to extract information on the original graph and a way to compare different structures of trading networks. The algorithm provides a direct way to distinguish their different hierarchies and structural organization. Each shell is labeled by a single shell index with the nodes colored uniquely. The nodes in the core are drawn in red. The size of a node is proportional to its original degree. The visualization used a logarithmic scale for the sizes in Fig.~\ref{Fig:STN_Kshell}. The maximum degree is 1261 in the core of the trading network. For clarity, we randomly select 5\% of edges and draw the nodes with the shell index $k\geq 5$ in Fig.~\ref{Fig:STN_Kshell}.

\section{Basic statistics of immediate price impact and trade size at the transaction level}

\begin{figure*}[htb]
  \centering
  \includegraphics[width=4.3cm]{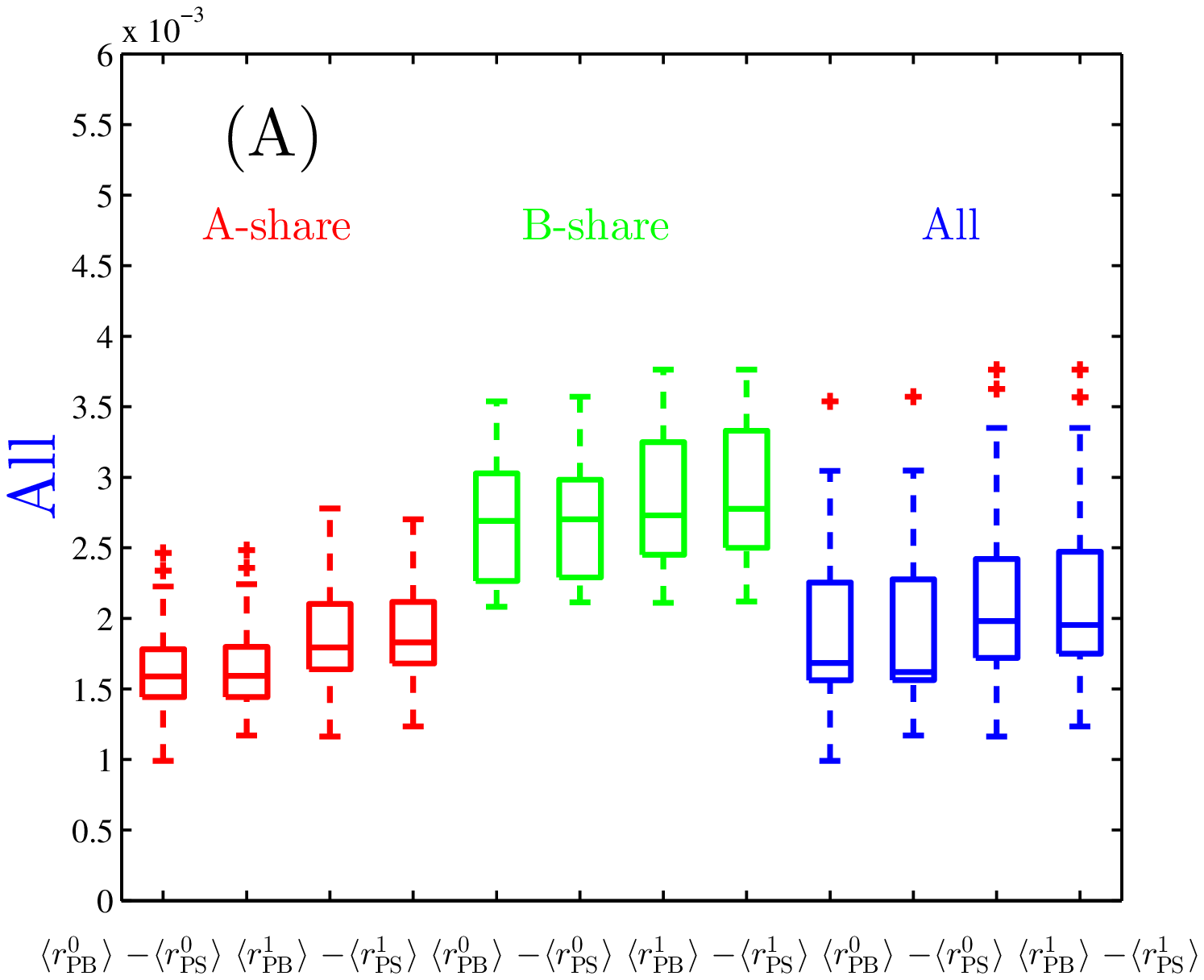}
  \includegraphics[width=4.3cm]{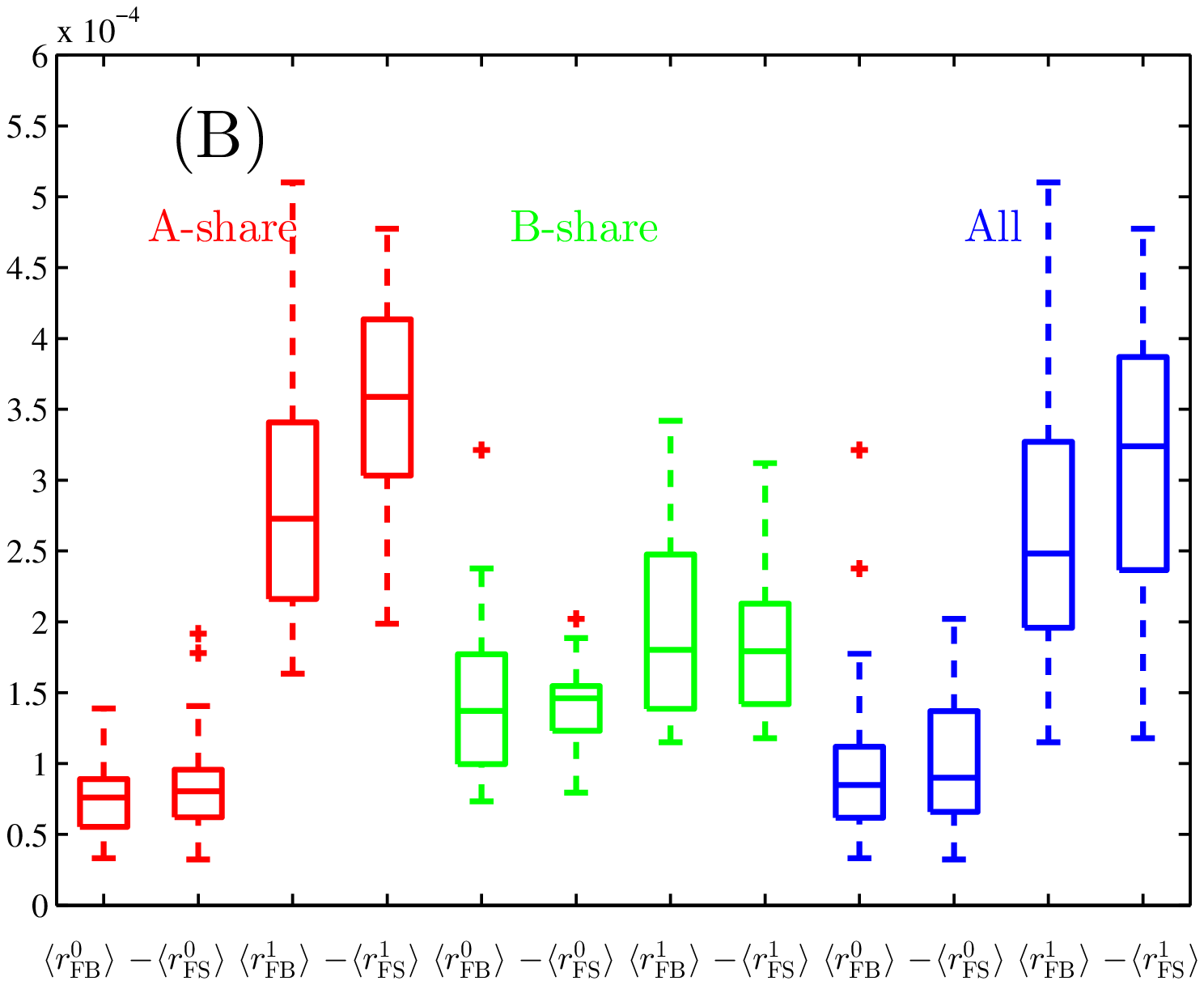}
  \includegraphics[width=4.3cm]{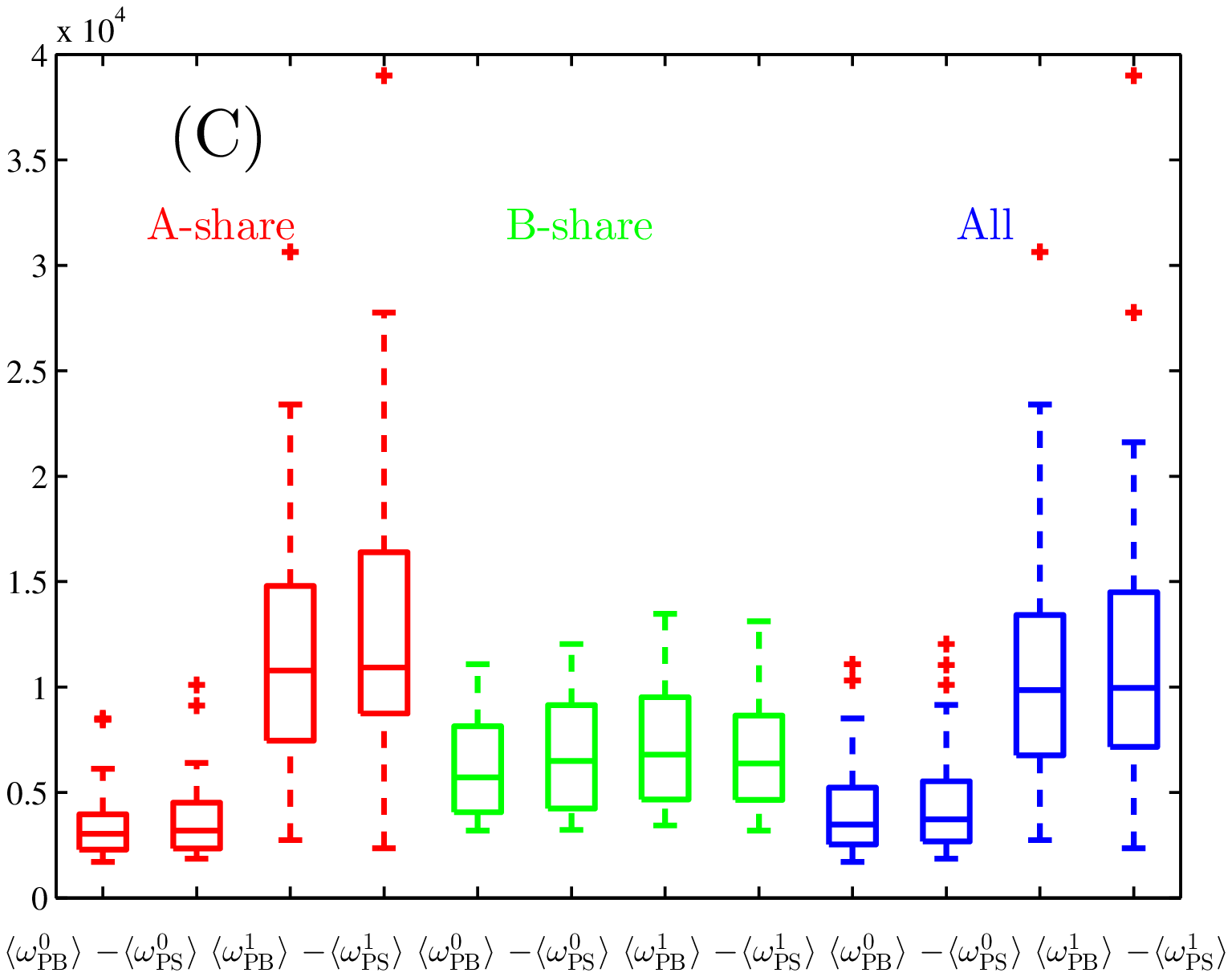}
  \includegraphics[width=4.3cm]{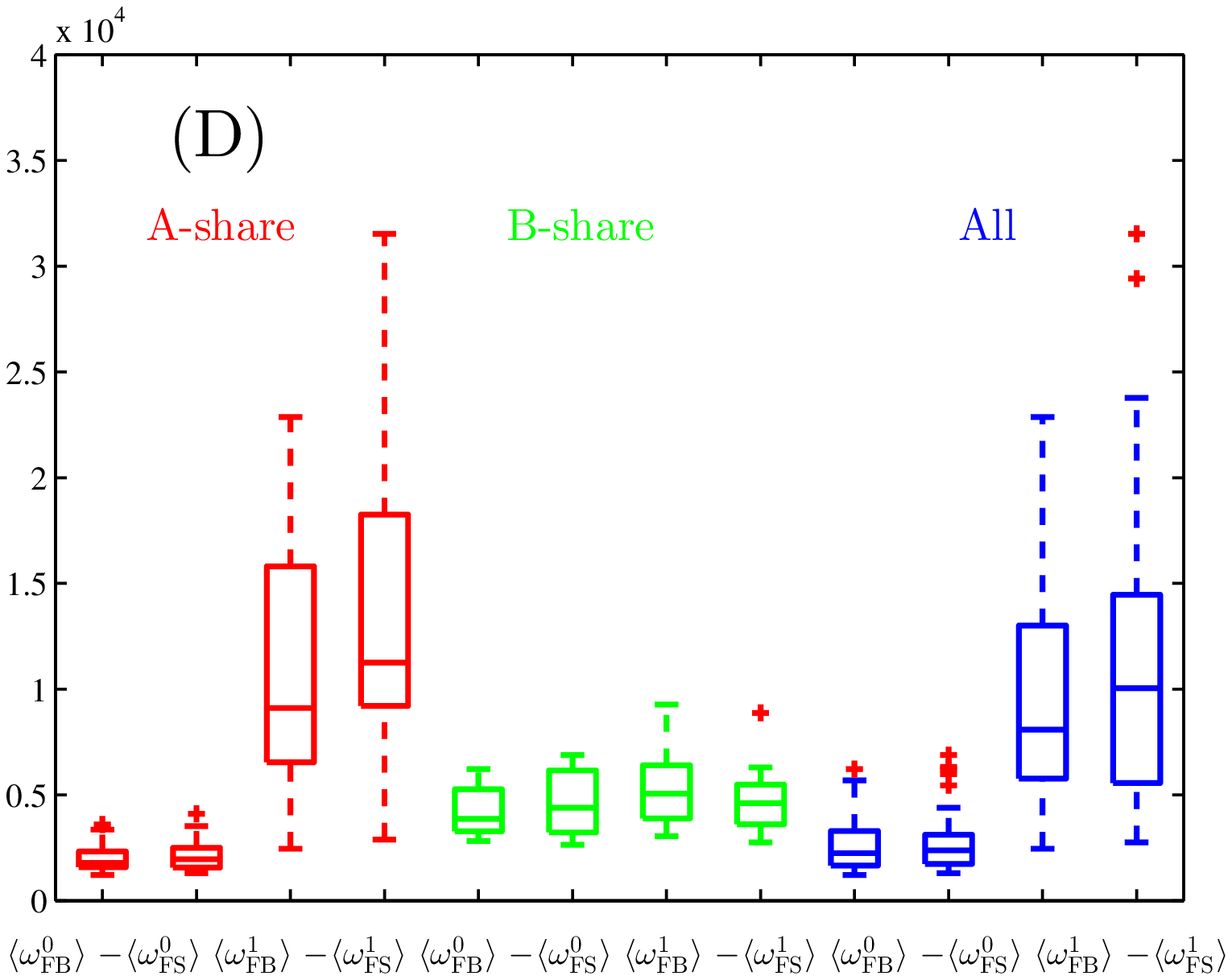}
  \includegraphics[width=4.3cm]{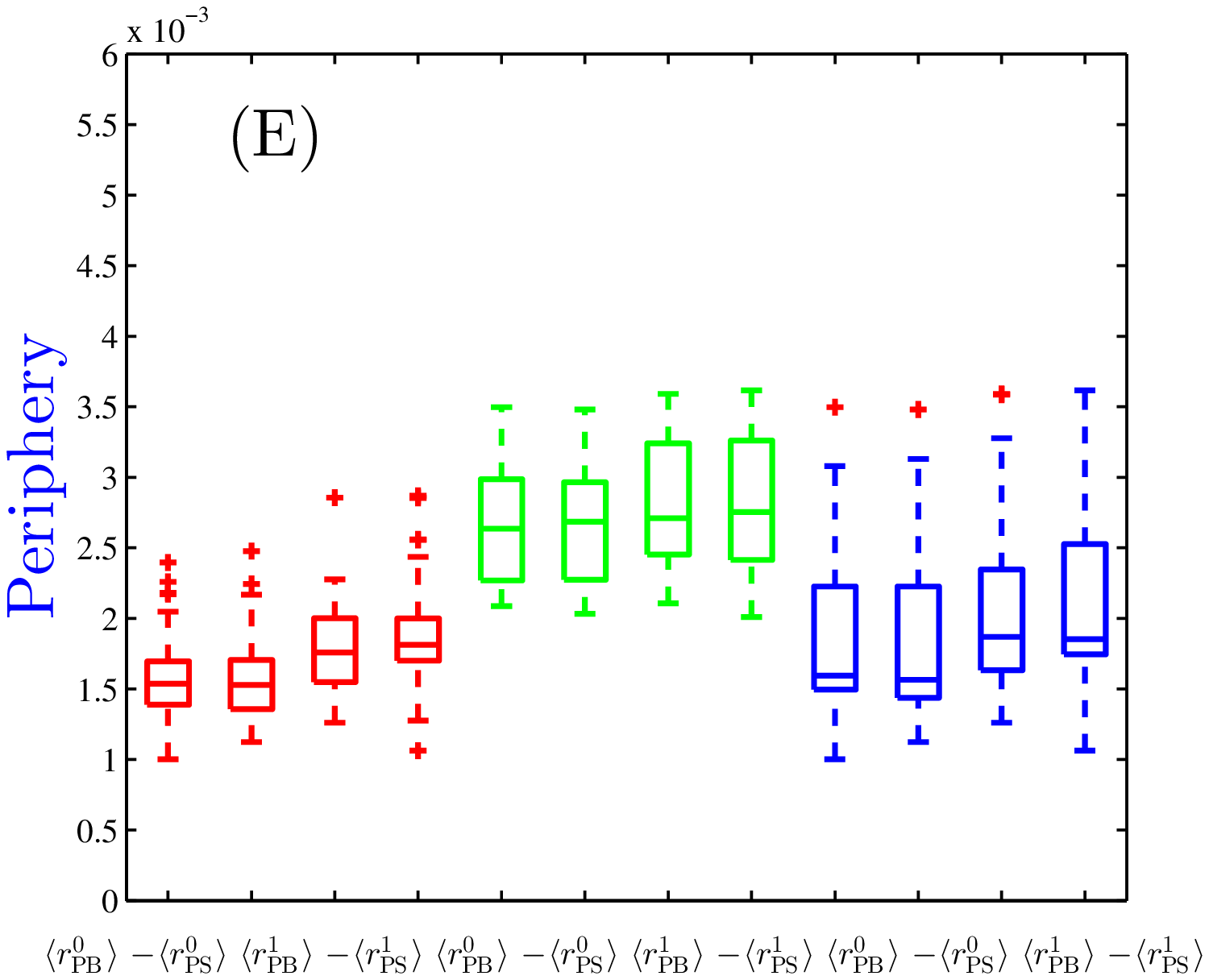}
  \includegraphics[width=4.3cm]{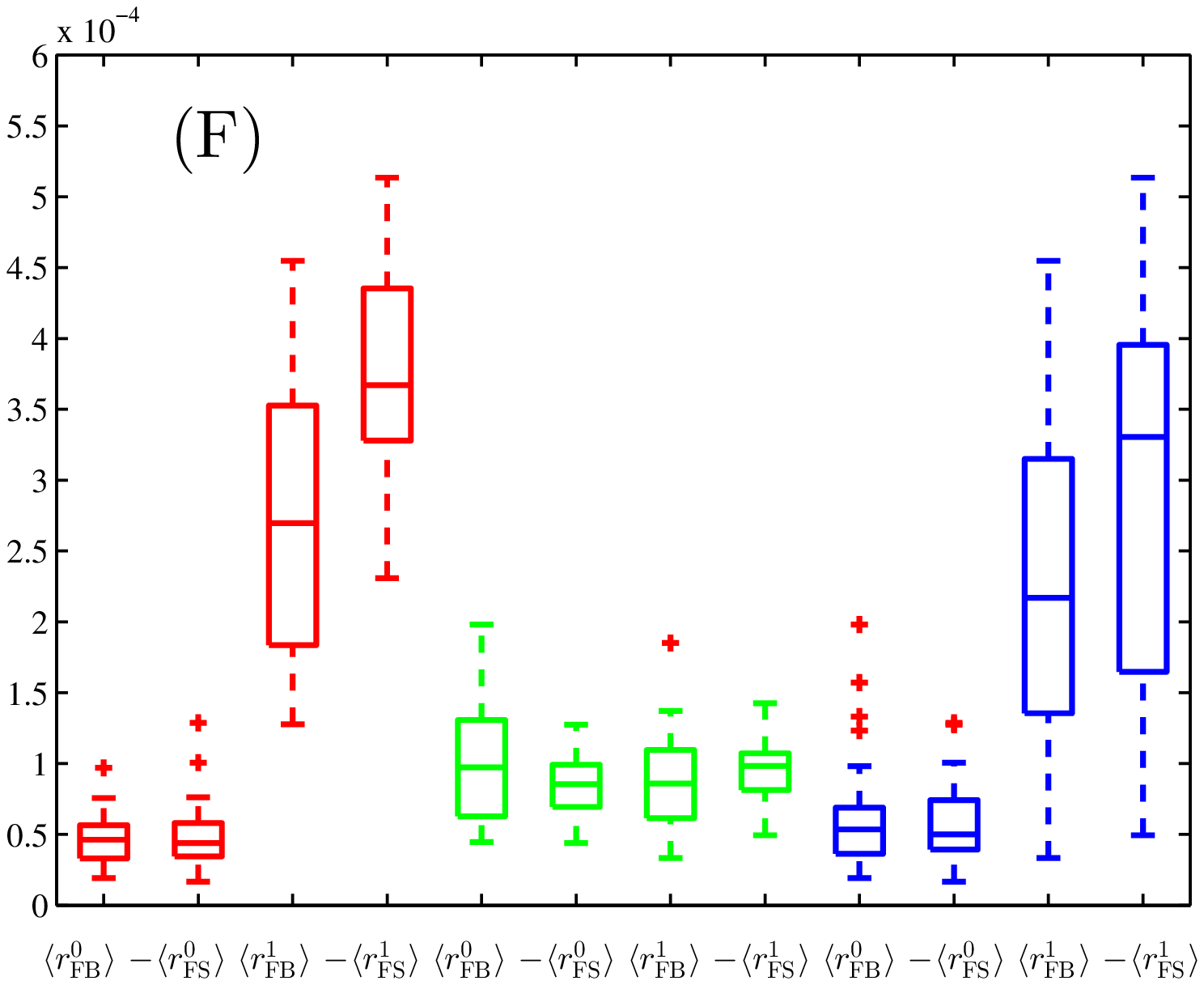}
  \includegraphics[width=4.3cm]{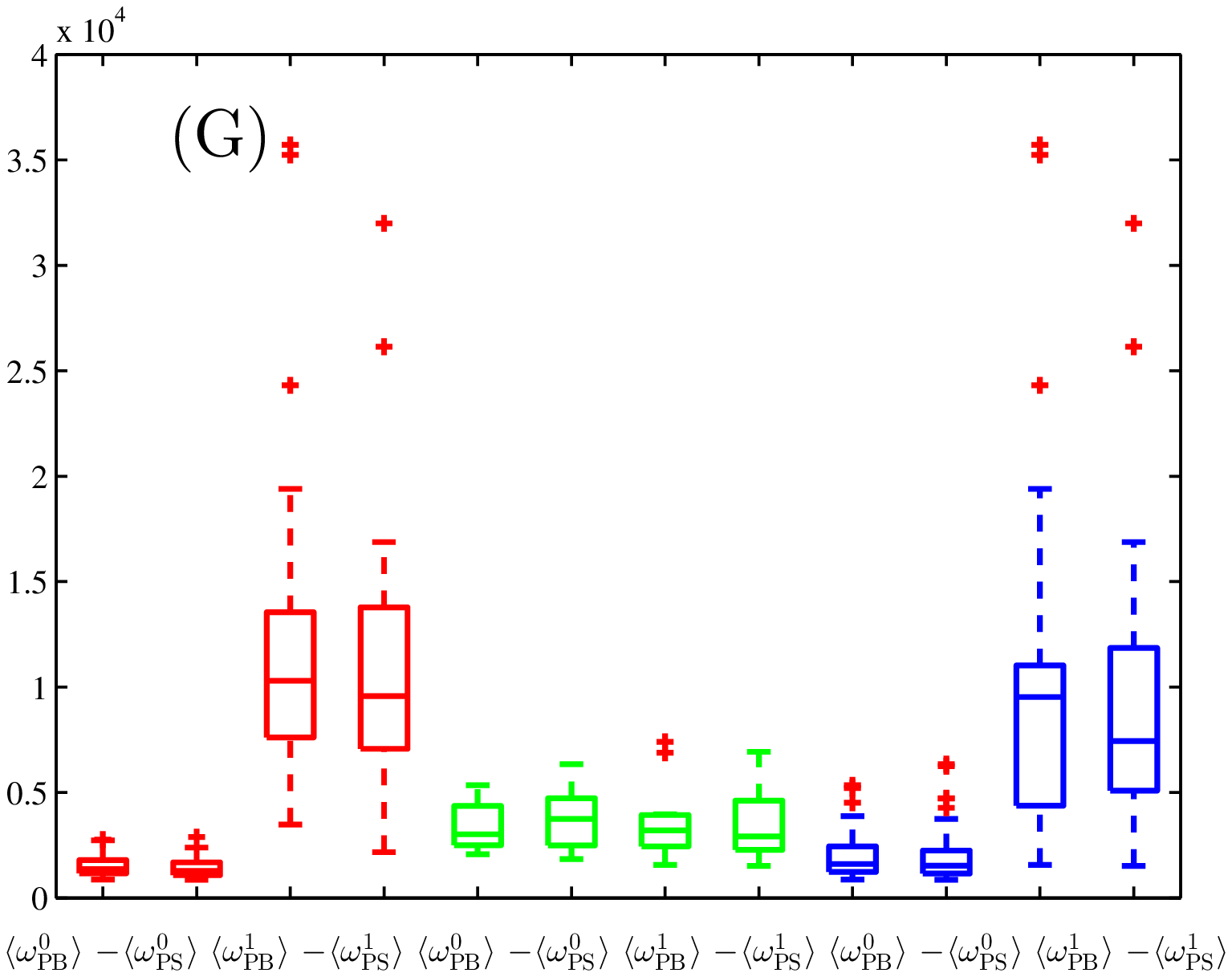}
  \includegraphics[width=4.3cm]{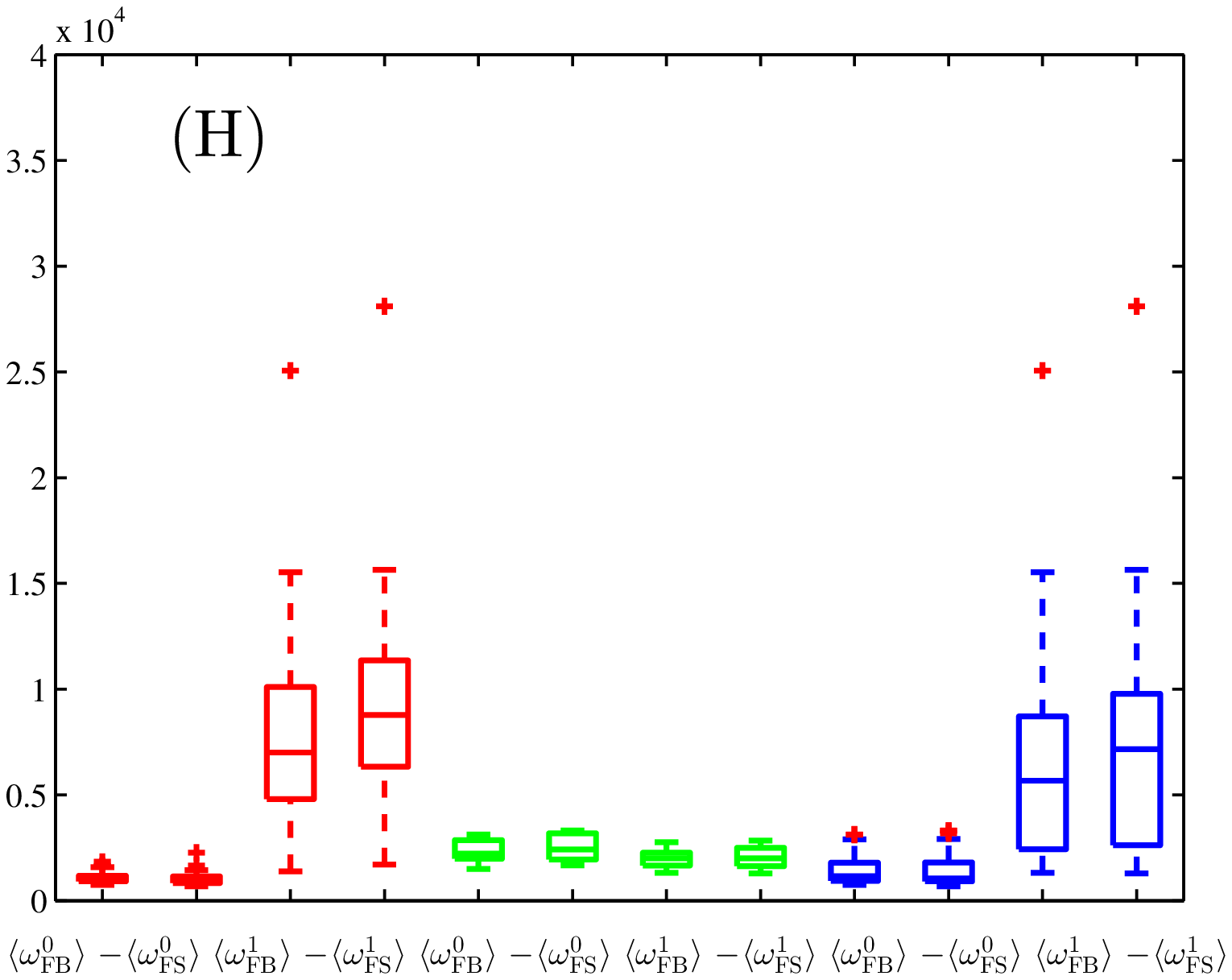}
  \includegraphics[width=4.3cm]{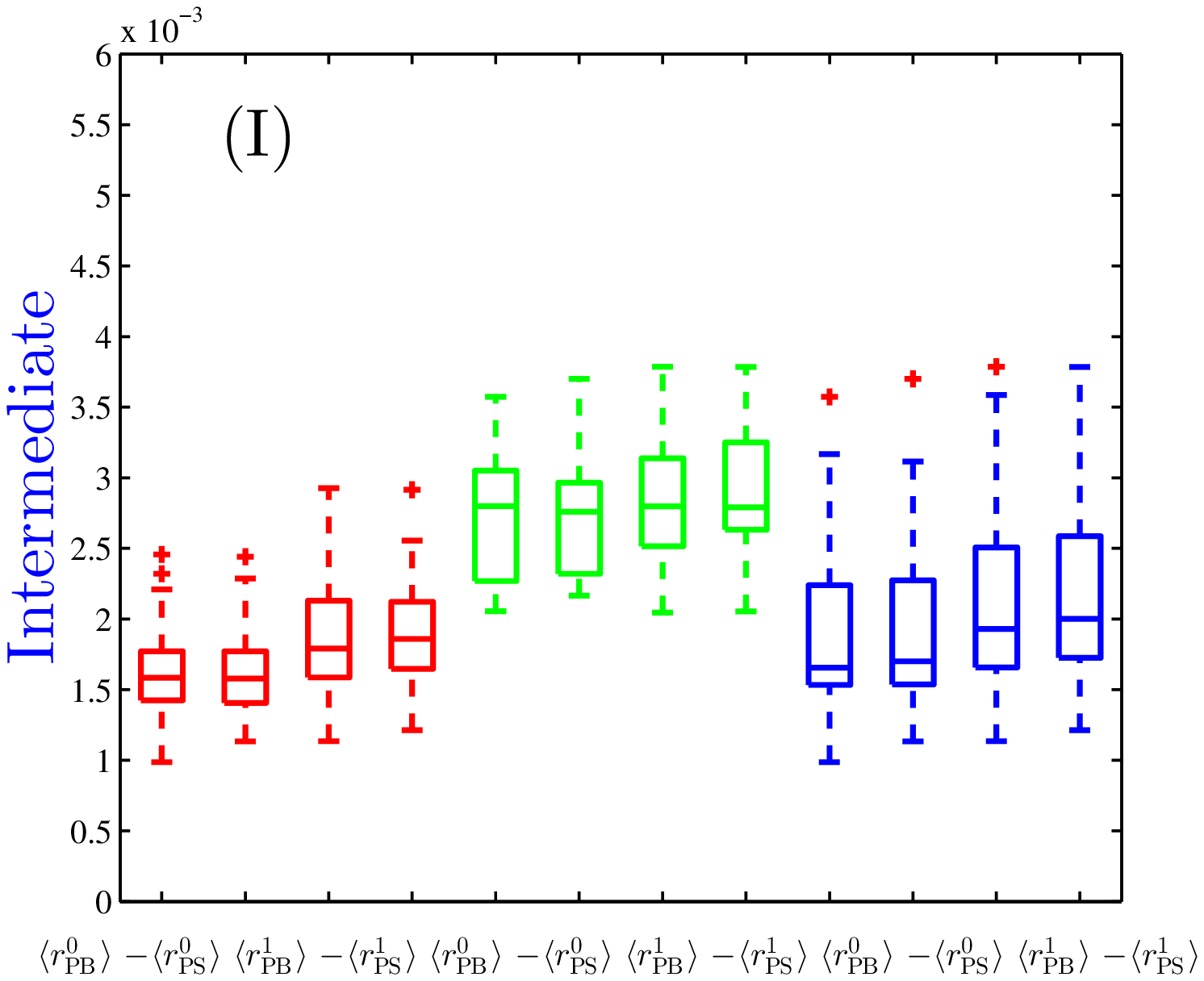}
  \includegraphics[width=4.3cm]{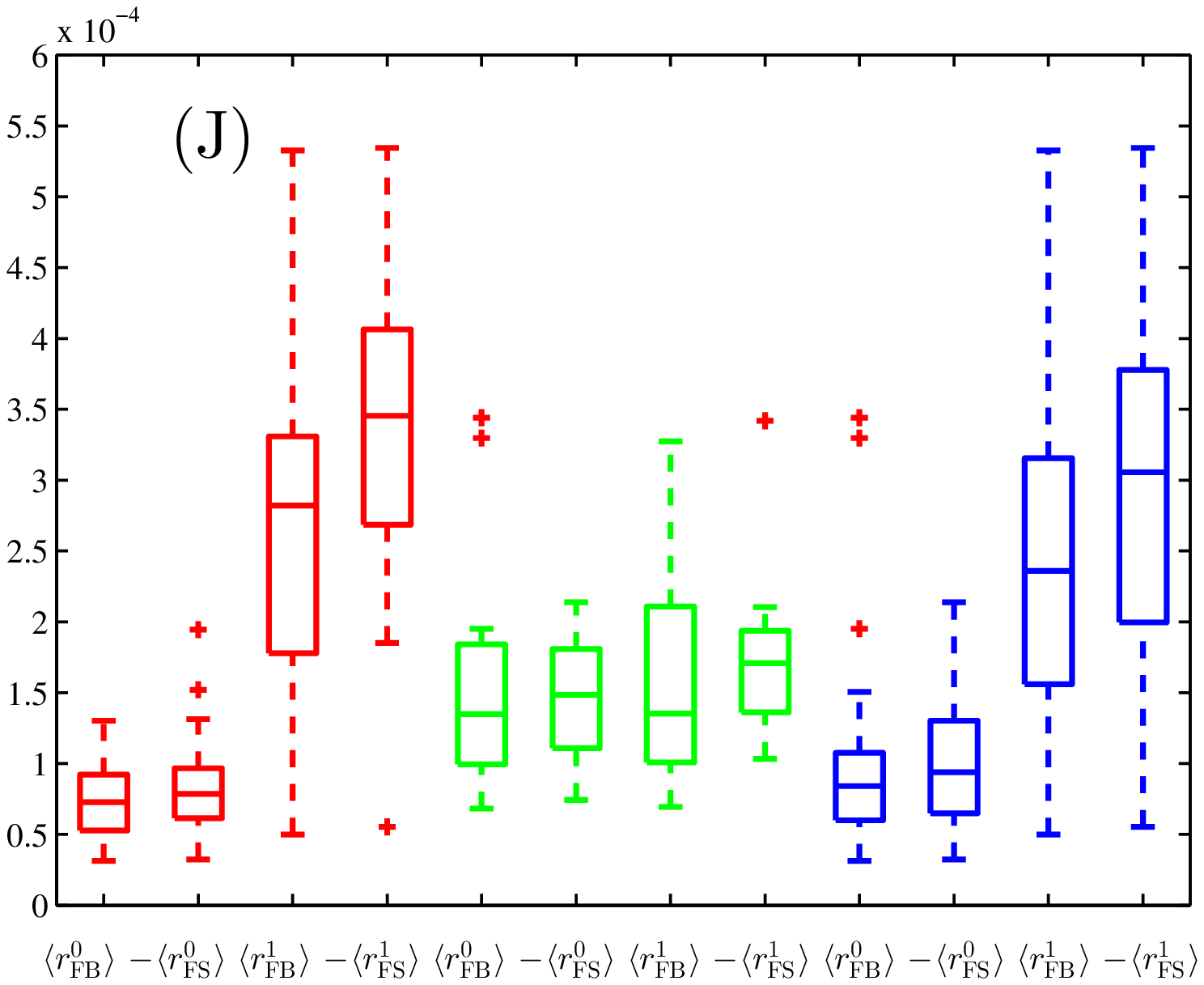}
  \includegraphics[width=4.3cm]{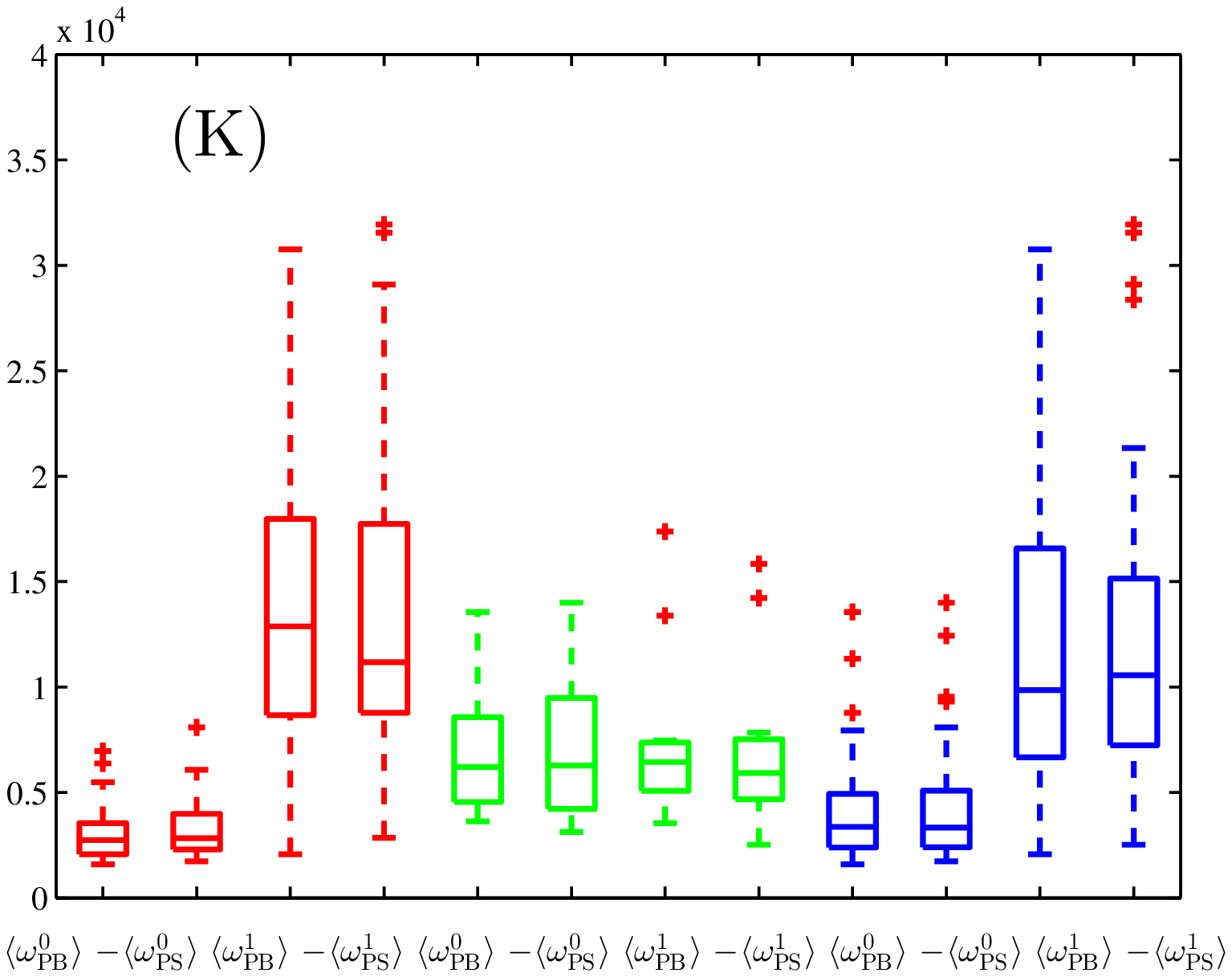}
  \includegraphics[width=4.3cm]{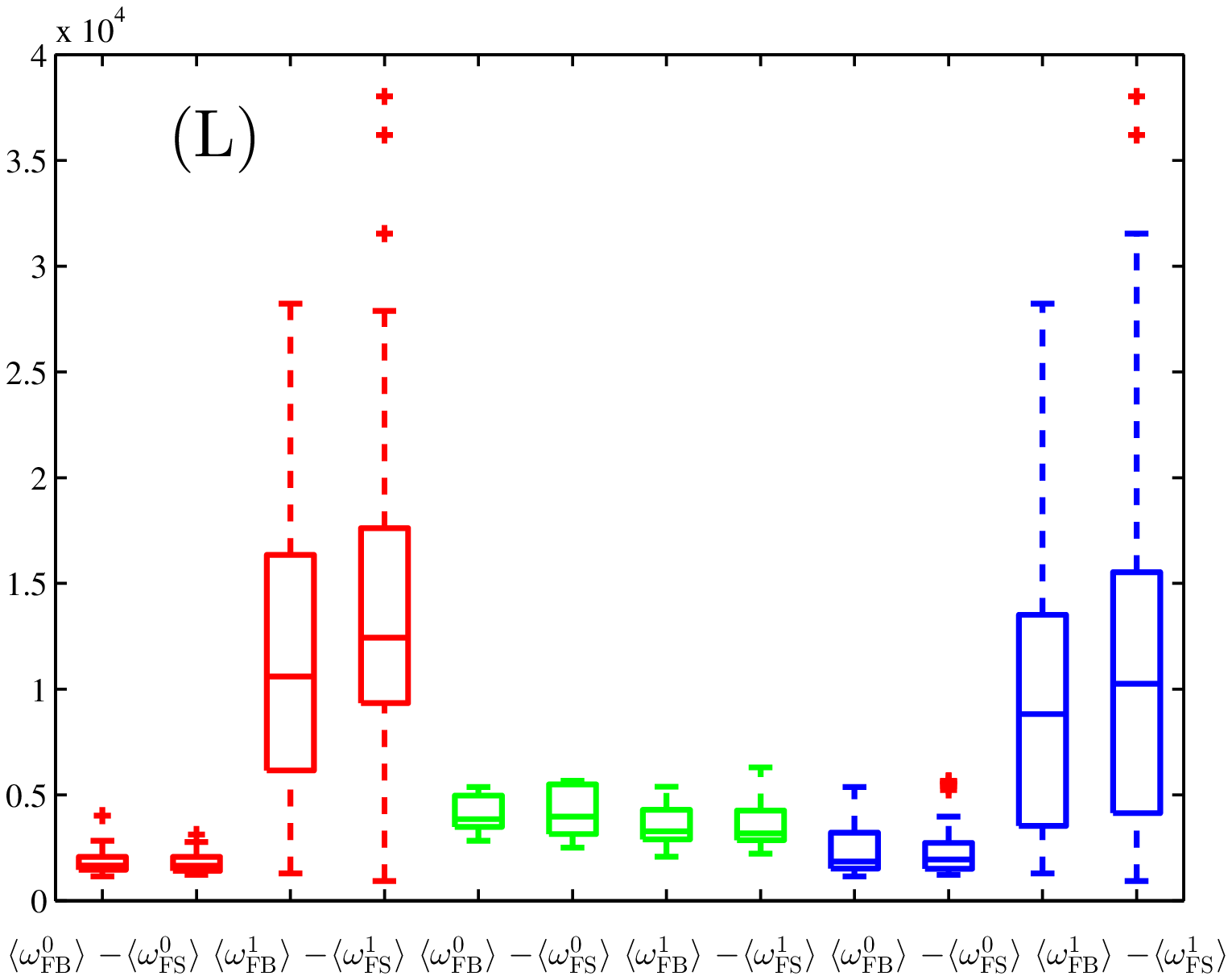}
  \includegraphics[width=4.3cm]{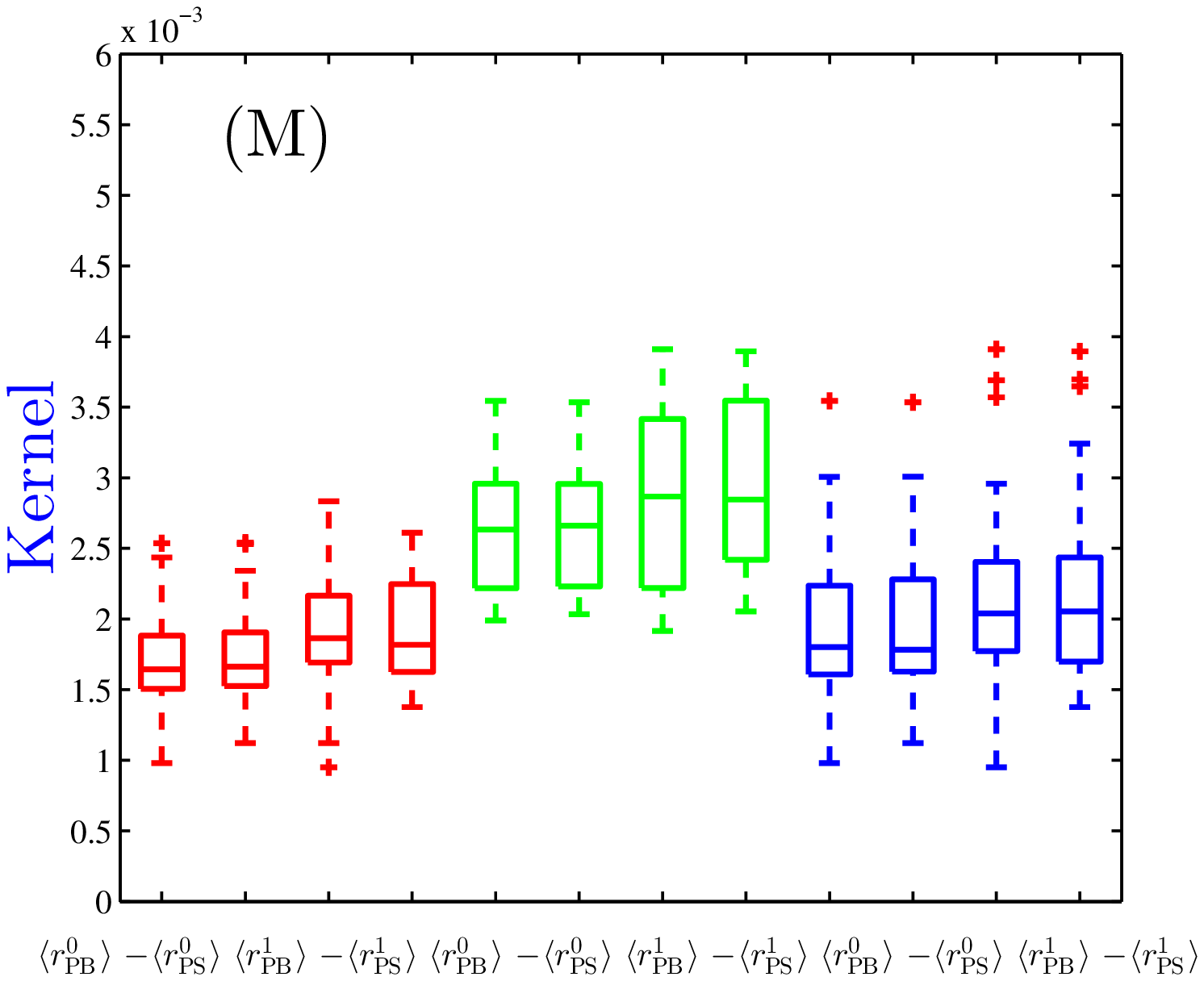}
  \includegraphics[width=4.3cm]{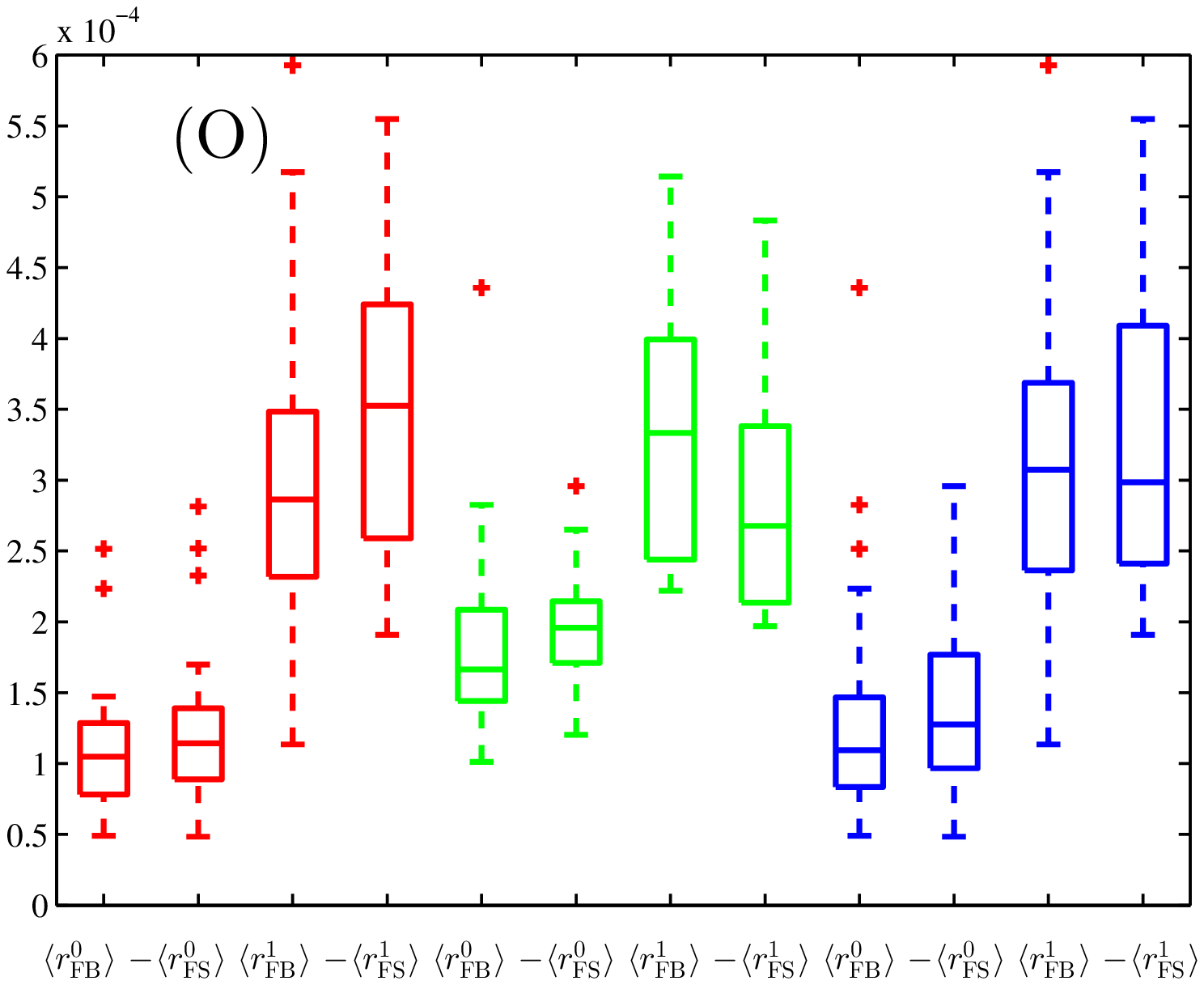}
  \includegraphics[width=4.3cm]{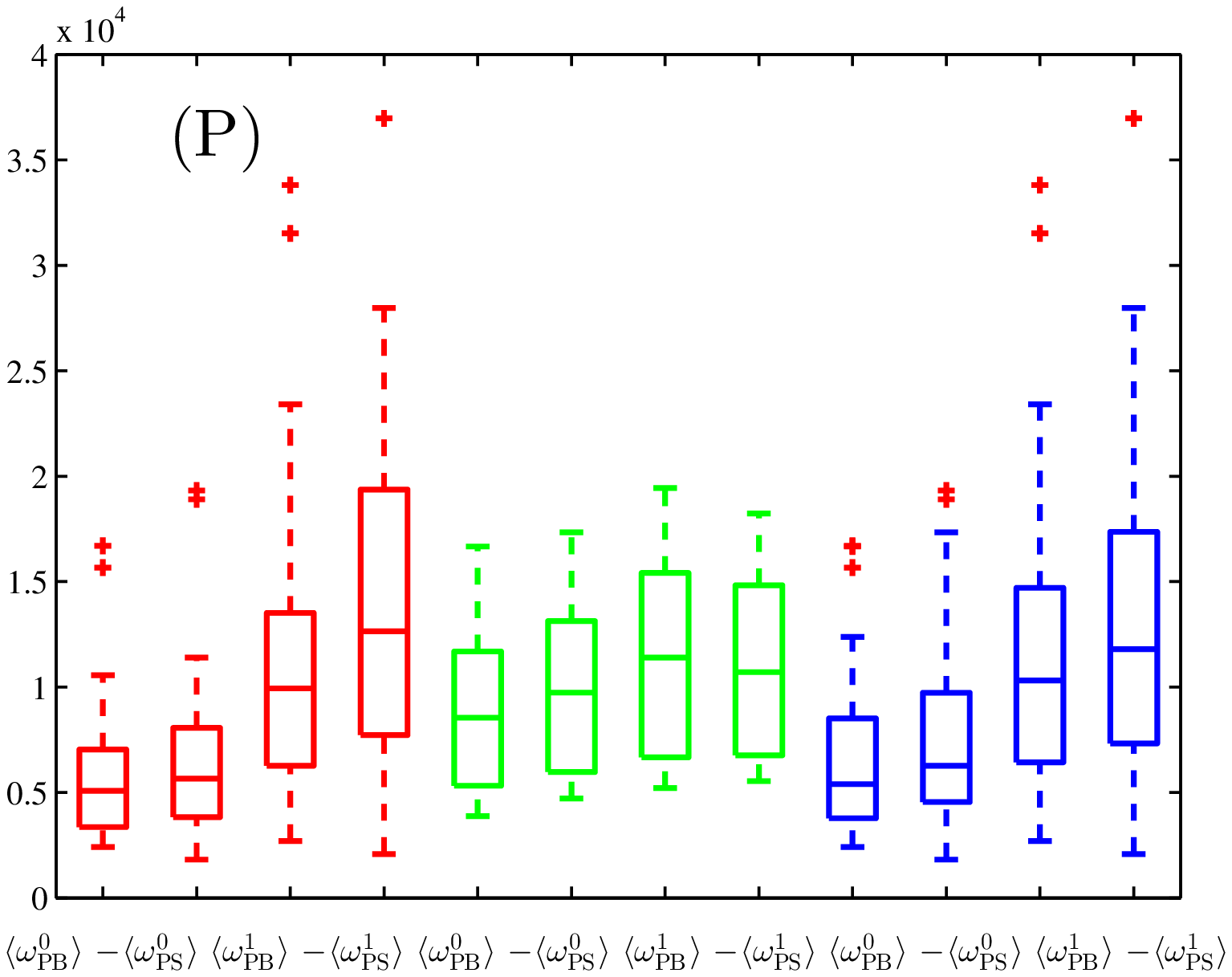}
  \includegraphics[width=4.3cm]{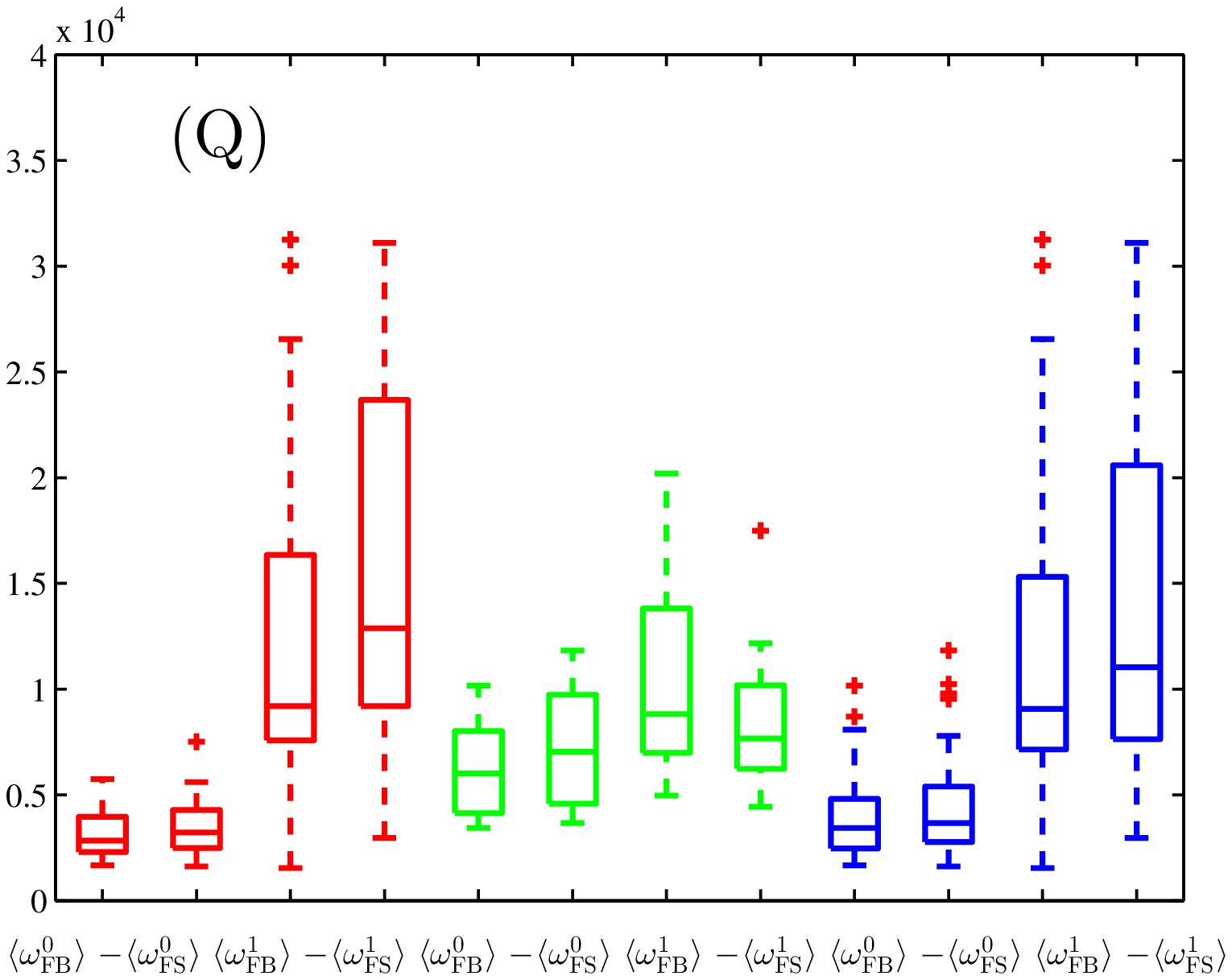}
  \caption{(color online) Box plots of average returns $\langle{r}\rangle$ and average trade sizes $\langle\omega\rangle$ at the transaction level for A-share market (red), B-share market (green), and all stocks (blue). For simplicity, we use subscripts PB, PS, FB, and FS to represent respectively the four types of trades. The superscripts 0 and 1 stand respectively for individuals and institutions. The four columns from left to right correspond to $\langle{r}\rangle$ of PB/PS trades, $\langle{r}\rangle$ of FB/FS trades, $\langle\omega\rangle$ of PB/PS trades, and $\langle\omega\rangle$ of FB/FS trades. The four rows from top to bottom correspond to all trades, periphery trades, intermediate trades, and kernel trades.}
  \label{Fig:IPI:omega:Stat}
\end{figure*}

The immediate price impact can be calculated as the percentage of mid-price change caused by a trade at time $t$
\begin{equation}
 r(t+1)=[p(t+1)-p(t)]/p(t)~,
 \label{Eq:rt}
\end{equation}
where $p(t)$ and $p(t+1)$ are the mid-prices of the best bid and ask right before and after the transaction at time $t$. The data set allows us to compare a closed national market (A-shares) with an international market (B-shares), individuals and institutions, partially filled and filled trades, buyer-initiated and seller-initiated trades. In this Letter, the two types of traders are individuals (superscript `0') and institutions (superscript `1'). The trades are divided into four types according to their directions and aggressiveness \cite{Biais-Hillion-Spatt-1995-JF,Zhou-2012-QF}: buyer-initiated partially filled (PB) trades resulting from partially filled buy orders, seller-initiated partially filled (PS) trades resulting from partially filled sell orders, buyer-initiated filled (FB) trades resulting from filled buy orders, and seller-initiated filled (FS) trades resulting from filled sell orders. As an example, $\omega_{\rm{FS}}^{0}$ stands for the transaction size of seller-initiated filled trades of individuals. The average return $\langle r\rangle$ and average transaction size $\langle \omega\rangle$ corresponding to the four types of trades are shown in the fist row of Fig.~\ref{Fig:IPI:omega:Stat}.

We classify the trades according to their positions in the trading networks by using the $k$-shell algorithm, which divides traders into $k_{\rm{max}}$ shells. For each stock trading network, the position of each trader in the network is denoted by the shell index $k\in \{1,2,...,k_{\rm{max}}\}$. The $k_{\rm{max}}$-shell is the core of trading network. The position of trades, initiated by traders in position $k$, is denoted by the corresponding shell index $k$. We sort the trades in ascending order depending on the value of the trades' position $k$ and divide trades into three equal parts, termed periphery, intermediate, and kernel. The periphery trades are initiated by traders in the peripheral shell and kernel trades are initiated by traders in the $k_{\max}$-core. We investigate whether traders at different network positions (periphery, intermediate and kernel) have different behaviors. The average return $\langle r\rangle$ and average transaction size $\langle \omega\rangle$ corresponding to the four types of trades at different network positions are shown in the second, third and fourth rows of Fig.~\ref{Fig:IPI:omega:Stat}.

According to the first and second columns of Fig.~\ref{Fig:IPI:omega:Stat}, the most intriguing feature is that the immediate price impact of partially filled trades is about 10 times of that of the filled trades \cite{Zhou-2012-QF}. The absolute immediate price impact of partially filled trades has an order of $10^{-3}$, while that of filled trades has an order of $10^{-4}$. Moreover, according to Fig.~\ref{Fig:IPI:omega:Stat}, there is no evident difference in the average sizes and the price impacts between buyer-initiated trades and seller-initiated trades.

\begin{figure*}[htb]
  \centering
  \includegraphics[width=8cm]{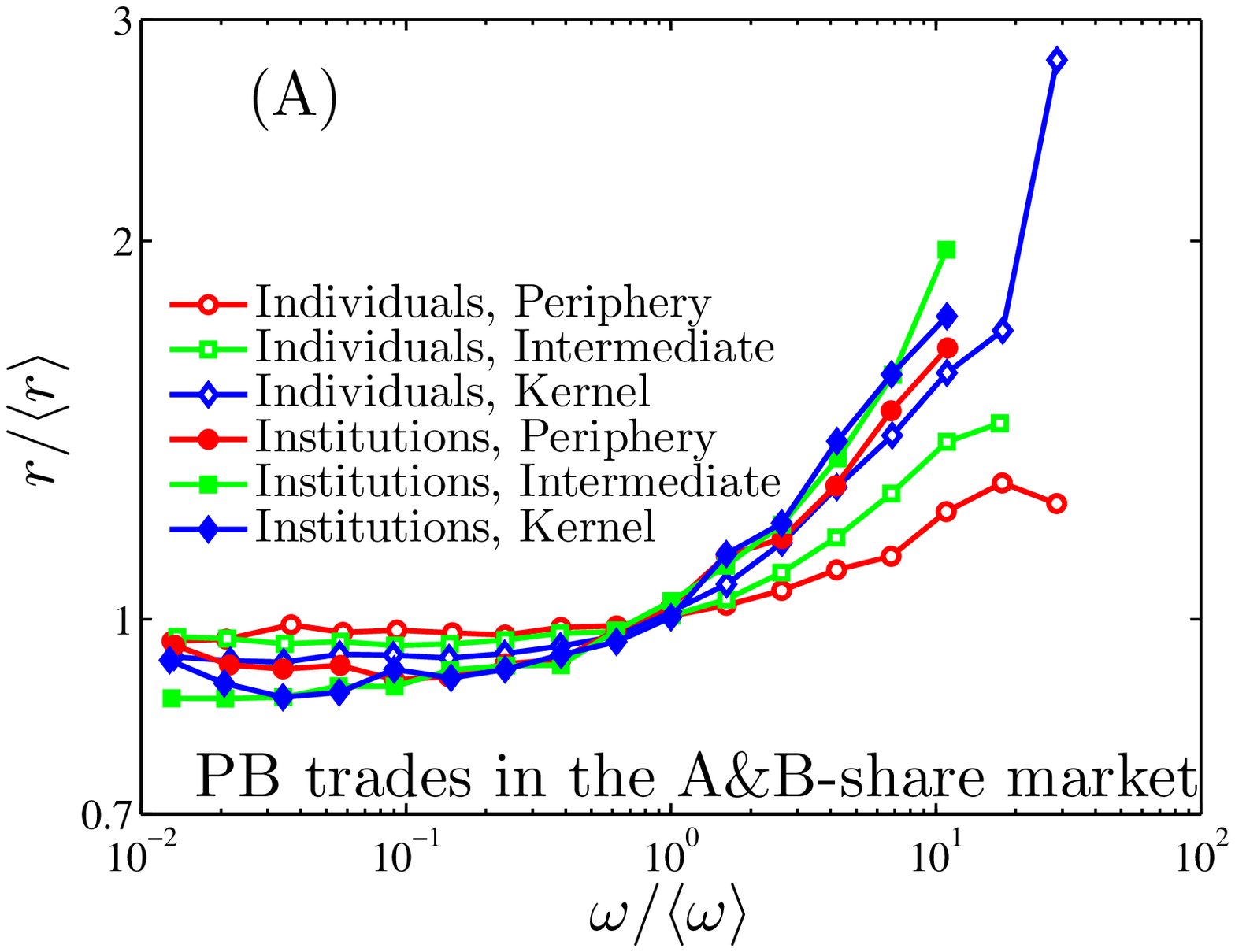}
  \includegraphics[width=8cm]{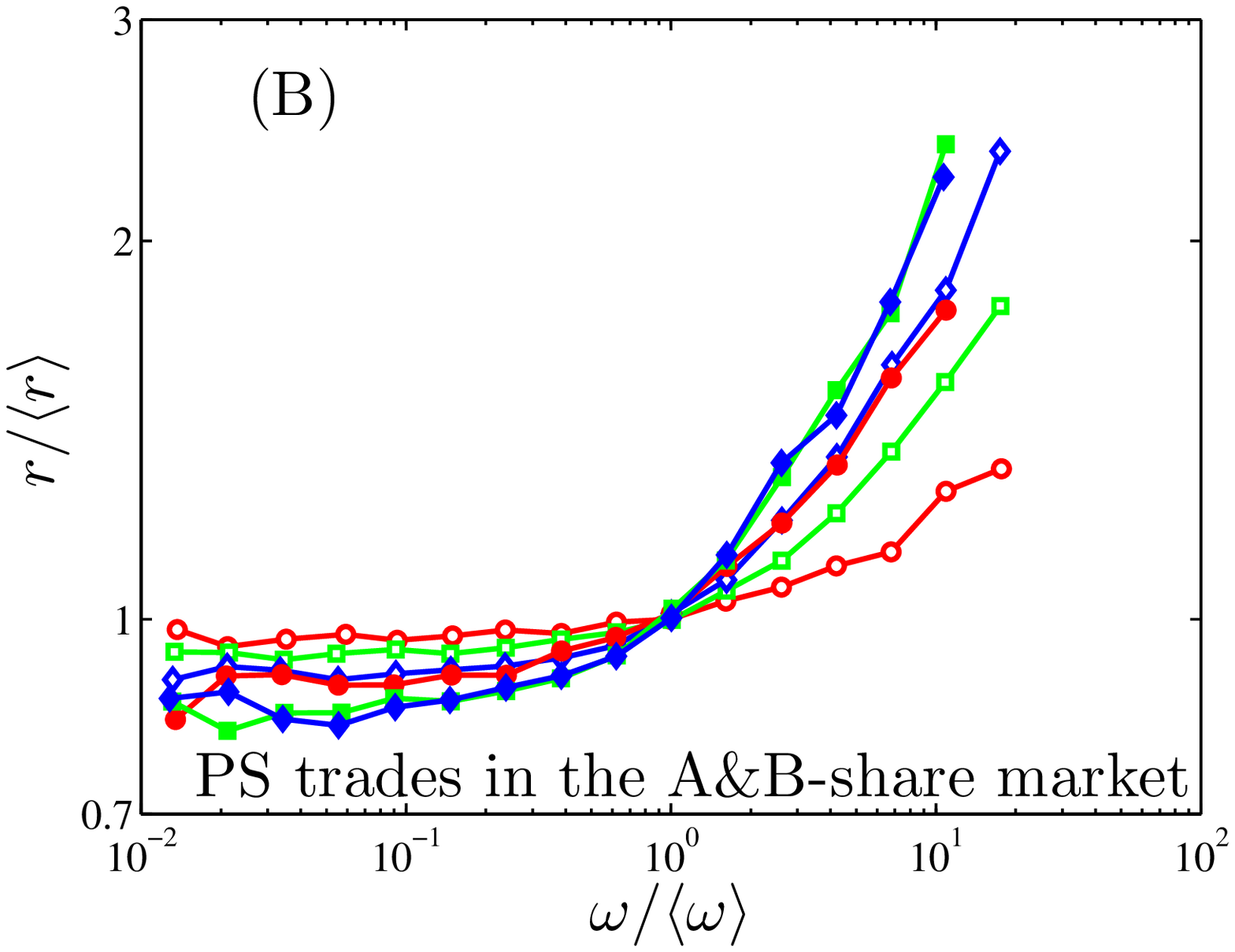}
  \includegraphics[width=8cm]{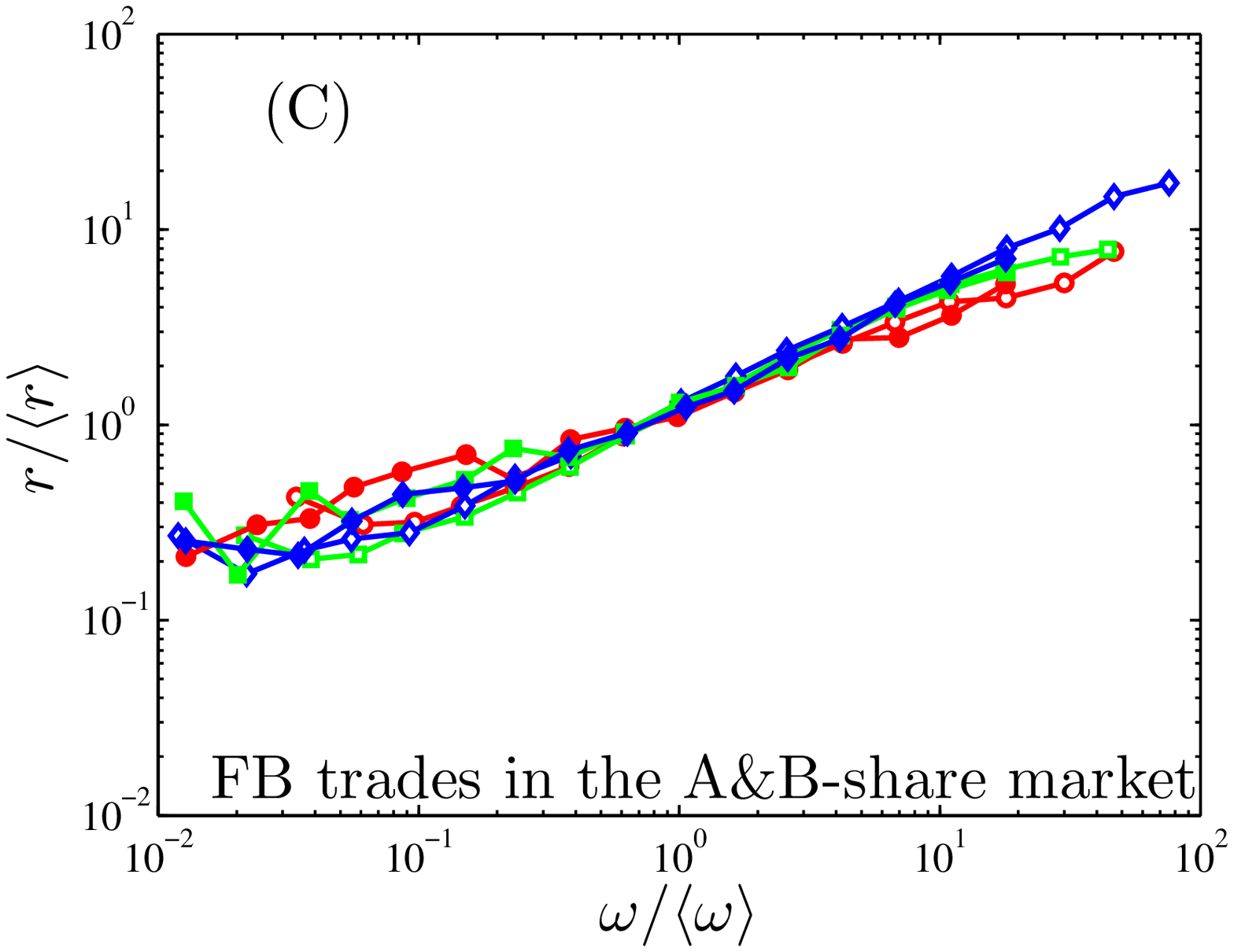}
  \includegraphics[width=8cm]{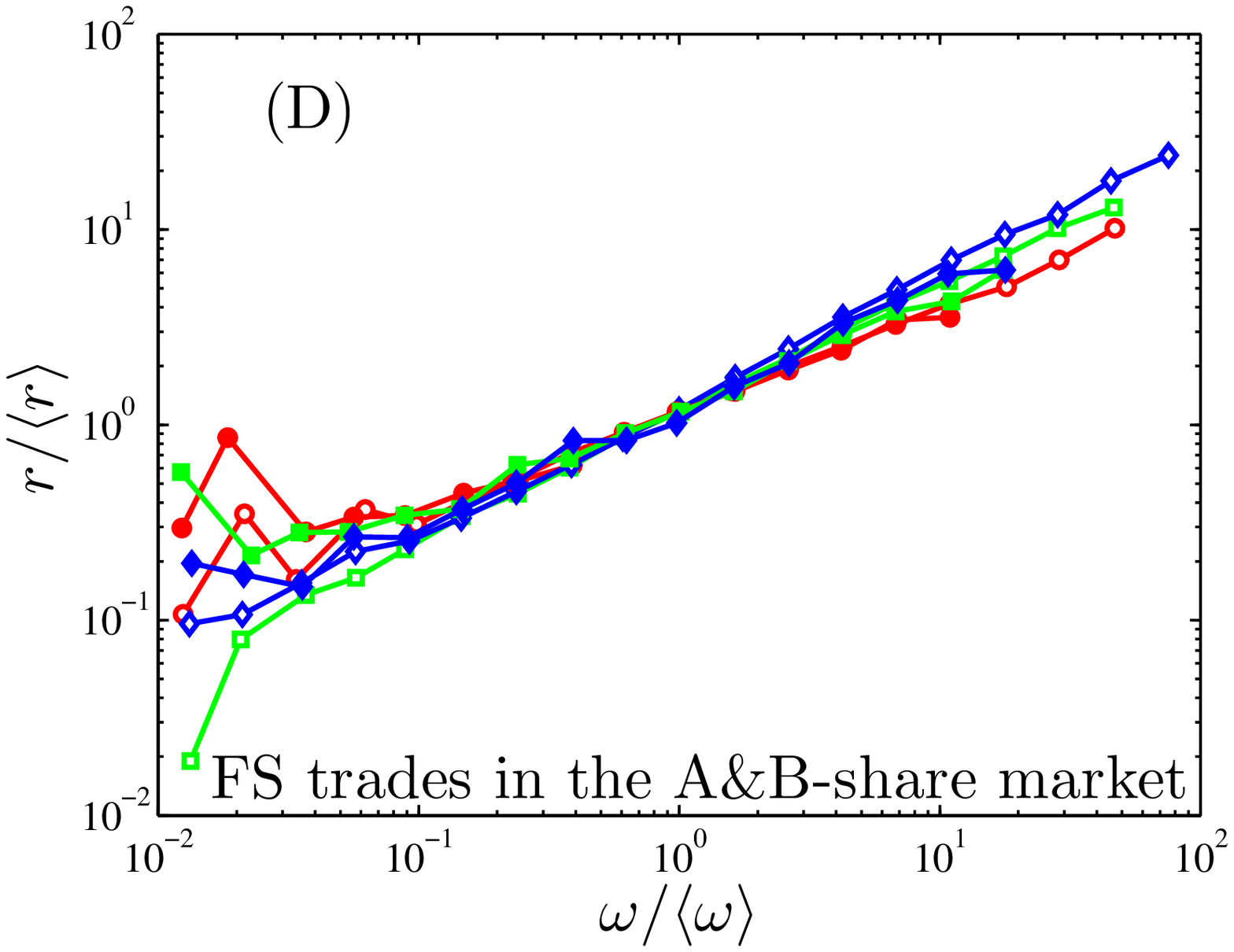}
  \caption{(color online) Normalized price impact of institutional and individual trades in both A-share and B-share markets at different positions of trading networks. The traders are classified into 3 types due to their positions (such as periphery, intermediate, and kernel) in the trading network for individuals and institutions. The results for the four types of trades due to their direction and aggressiveness are present in four plots: (A) Buyer-initiated partially filled trades, (B) Seller-initiated partially filled trades, (C) Buyer-initiated filled trades, and (D) Seller-initiated filled trades.}
  \label{Fig:BestBidAsk_Impact01}
\end{figure*}

We find that the average trade size of each type (PB, PS, FB, and FS) submitted by individual traders in the B-share market is larger than that in the A-share market, thus the absolute price impact is also larger for B-share trades. In contrast, the average size of B-share trades is smaller than A-share trades for institutions. It is reasonable that the absolute price impact of filled A-share trades submitted by institutions is larger than filled B-share trades, but partially filled A-share trades have smaller absolute price impact than partially filled B-share trades. This abnormal phenomenon cannot be explained by the price gaps in the two markets \cite{Zhou-2012-QF,Zhou-2012-NJP,Gu-Xiong-Zhang-Chen-Zhang-Zhou-2016-CSF}.

We observe, for each of the four types of trades (PB, PS, FB, and FS), institutional traders have larger immediate price impacts than individual traders do:
\begin{equation}
  \langle{r}^1_{\rm{PB,FB}}\rangle > \langle{r}^0_{\rm{PB,FB}}\rangle ~~{\rm{and}}~~
  -\langle{r}^1_{\rm{PS,FS}}\rangle > -\langle{r}^0_{\rm{PS,FS}}\rangle,
  \label{Eq:r1:r0}
\end{equation}
which is mainly caused by the fact that institutional traders submit larger orders than individual traders:
\begin{equation}
  \langle{\omega}^1_{\rm{PB,PS,FB,FS}}\rangle > \langle{\omega}^0_{\rm{PB,PS,FB,FS}}\rangle.
  \label{Eq:w1:w0}
\end{equation}
This phenomenon is particularly evident for A-share stocks and is marginal for B-share stocks.

In most cases, we find that trades closer to the kernel have larger sizes and high immediate price impacts:
\begin{equation}
  \left\{
  \begin{array}{lll}
   \begin{aligned}
   & \langle{r}_{\rm{K}}\rangle > \langle{r}_{\rm{I}}\rangle > \langle{r}_{\rm{P}}\rangle\\
   & \langle{\omega}_{\rm{K}}\rangle > \langle{\omega}_{\rm{I}}\rangle > \langle{\omega}_{\rm{P}}\rangle
     \end{aligned}
  \end{array}
  \right..
  \label{Eq:r:w:P:I:K}
\end{equation}
This observation simply indicates that large trades are more likely to be executed with more other trades and incur larger price impacts.

\section{Relationship between immediate price impact and trade size at different network positions}

We normalize $r$ and $\omega$ for each type of trades in both A-share and B-share markets by their averages $\langle{r}\rangle$ and $\langle\omega\rangle$ \cite{Zhou-2012-QF}. For partially filled trades in Fig.~\ref{Fig:BestBidAsk_Impact01} (A) and (B), when $\omega$ is smaller than the average $\langle\omega\rangle$, the price impact $r(\omega)$ is not sensitive to trade size $\omega$, and the normalized price impact of individuals is slightly larger than that of institutions in each position. Furthermore, the normalized price impact is slightly larger for traders with small $k$. When $\omega>\langle\omega\rangle$, $r(\omega)$ exhibits a significant upward trend, the normalized price impact of individuals is smaller than that of institutions in each position, and $r(\omega)/\langle{r}\rangle$ is larger for traders with large $k$.

For filled trades in Fig.~\ref{Fig:BestBidAsk_Impact01} (C) and (D), we can observe power-law scaling behaviours between the normalize $r$ and $\omega$ for different types of trades in different positions:
\begin{equation}\label{Eq:r:w}
  r(\omega)/\langle{r}\rangle \sim (\omega/\langle\omega\rangle)^\alpha,
\end{equation}
where $\alpha$ is the power-law scaling exponent. The power-law scaling range spans about three orders of magnitude from $10^{-1}$ to $10^{2}$. These two plots extend precious results for which the power-law dependence of the price impact on the trade size holds for FB and FS trades \cite{Zhou-2012-QF}, such that this law holds for individuals and institutions in different positions of trading networks.

\begin{table}[tb]
  \centering
  \caption{\label{Tb:alphas} Exponent $\alpha$ of power-law behaviour $r\sim\omega^{\alpha}$ for buyer-initiated filled trades and seller-initiated filled trades. The analysis is conducted respectively for all stocks, A-shares and B-shares. Due to their positions in the trading network, we divide the traders into 3 types for individuals and institutions, including periphery (P), intermediate (I) and kernel (K).}
\medskip
\begin{tabular}{cccccccccccc}
 \hline
Market  &  Type  &   Pos. & Individual & Institution  \\    \hline
  A \& B  &FB  & P  & $0.52\pm0.02$  & $0.46\pm0.03$  \\
          &FB  & I  & $0.58\pm0.02$  & $0.52\pm0.02$  \\
          &FB  & K  & $0.62\pm0.01$  & $0.58\pm0.02$  \\
          &FS  & P  & $0.55\pm0.01$  & $0.51\pm0.01$  \\
          &FS  & I  & $0.64\pm0.01$  & $0.57\pm0.02$  \\
          &FS  & K  & $0.69\pm0.01$  & $0.62\pm0.02$  \\
       \hline
  A-share &FB  & P  & $0.52\pm0.02$  & $0.39\pm0.03$  \\
          &FB  & I  & $0.58\pm0.02$  & $0.46\pm0.03$  \\
          &FB  & K  & $0.62\pm0.01$  & $0.52\pm0.01$  \\
          &FS  & P  & $0.55\pm0.01$  & $0.43\pm0.02$  \\
          &FS  & I  & $0.64\pm0.01$  & $0.53\pm0.02$  \\
          &FS  & K  & $0.69\pm0.00$  & $0.53\pm0.03$  \\
    \hline
  B-share &FB  & P  & $0.52\pm0.02$  & $0.58\pm0.03$  \\
          &FB  & I  & $0.65\pm0.02$  & $0.63\pm0.02$  \\
          &FB  & K  & $0.62\pm0.03$  & $0.69\pm0.02$  \\
          &FS  & P  & $0.62\pm0.01$  & $0.58\pm0.03$  \\
          &FS  & I  & $0.71\pm0.02$  & $0.59\pm0.04$  \\
          &FS  & K  & $0.70\pm0.03$  & $0.73\pm0.03$  \\
  \hline
 \end{tabular}
\end{table}

The power-law scaling exponents $\alpha$ are obtained by linear regressions of $\ln[r(\omega)/\langle{r}\rangle]$ against $\ln[\omega/\langle\omega\rangle]$, which are presented in table \ref{Tb:alphas}. A comparison of individual and institutional trades in the same position shows that
\begin{equation}
 \alpha^{0}>\alpha^{1}.
 \label{Eq:alpha:0:1}
\end{equation}
It means that the price impact of filled trades of individuals is much more sensitive to the trade size than that of institutions. Generally, institutions are more professional than individuals in financial markets. To reduce transaction costs and risks, institutional traders use certain trading strategies to reduce their price impact. In contrast, most of individual traders place orders with worse strategies \cite{Zhou-Mu-Chen-Sornette-2011-PLoS1,Zhou-Mu-Kertesz-2012-NJP}. A comparison of trades in different positions shows that
the power-law exponents $\alpha$ for trades in different positions of the trading networks have the following relationship:
\begin{equation}
  \alpha_{\rm{P}}<\alpha_{\rm{I}}<\alpha_{\rm{K}}.
  \label{Eq:alpha:P:I:K}
\end{equation}
The price impact of filled kernel trades is more sensitive to the trade size than the trades outer shells. Moreover, we observe certain asymmetry between buy trades and sell trades such that
\begin{equation}
 \alpha^{\rm{FS}}>\alpha^{\rm{FB}}.
 \label{Eq:alpha:FS:FB}
\end{equation}
This buy/sell asymmetry was not observed when one did not look into trade positions \cite{Zhou-2012-QF}.

We perform the same analysis on A-share stocks and B-share stocks separately. Nice power-law dependence of the price impact on the trade size is also observed. The power-law scaling exponents are presented in table~\ref{Tb:alphas}. We find that the relationships in Eq.~(\ref{Eq:alpha:P:I:K}) and Eq.~(\ref{Eq:alpha:FS:FB}) also hold for both individuals and institutions. The relationship in Eq.~(\ref{Eq:alpha:0:1}) holds for A-share stocks, but not for B-share stocks. Comparing the results of A-share stocks and B-share stocks, we find that
\begin{equation}
  \alpha^{\rm{B}}>\alpha^{\rm{A}},
  \label{Eq:alpha:A:B}
\end{equation}
where $\alpha^{\rm{A}}$ and $\alpha^{\rm{B}}$ stand respectively for the power-law exponents of A-share stocks and B-share stocks.

\section{Summary}

In this Letter, we have analyzed a large data set of order flows recorded in the Shenzhen Stock Exchange, focusing on the immediate price impact of institutional and individual trades in different positions of stock trading networks. We perform a statistical analysis of immediate price impact of all the traders trading 32 A-share stocks and 11 B-share stocks in 2003. The data offer a unique opportunity to compare a closed national market (A-shares) with an international market (B-shares), individuals and institutions, partially filled and filled trades, buyer-initiated and seller-initiated trades, and trades at different positions (periphery, intermediate and kernel) of trading networks.

We constructed the stock trading networks based on limit order book data and classified the traders into $k$ shells using the $k$-shell decomposition algorithm. Based on PB, PS, FB, and FS trades, we investigate different trading behaviors of individuals and institutions at different positions of trading network. Individual traders' filled trades are found to be more aggressive than institutional traders' filled trades. From periphery to intermediate to kernel, the immediate price impact of institutional and individual trades increases. The analysis has also been conducted separately for A-shares and B-shares and shows that trades in the B-shares market are more aggressive than in the A-shares market.

For filled trades, we confirmed the presence of power-law price impacts, which holds for trades with directions and aggressiveness, trades submitted by individuals and institutions, and trades at different positions of stock trading networks. Our findings thus extend previous results \cite{Zhou-2012-QF}. The main contribution of this Letter stems from the fact that we considered the topological structure of trading networks and used the information extracted from network positions, which has not been studies in the literature
\cite{Kraus-Stoll-1972-JF,Dufour-Engle-2000-JF,Saar-2001-RFS,Lillo-Farmer-Mantegna-2003-Nature,Chiyachantana-Jain-Jiang-Wood-2004-JF,Lim-Coggins-2005-QF,Zhou-2012-QF}.  Our findings shed new lights on interdisciplinary network analysis about topological structure and price impacts in complex trading networks.

\acknowledgments

This work is supported by the National Natural Science Foundation of China (71571121, 71501072, 71131007), the Postdoctoral Science Foundation of China (2015M571502, 2015M570342), and the Fundamental Research Funds for the Central Universities.


\begin{thebibliography}{10}
\expandafter\ifx\csname url\endcsname\relax\def\url#1{\texttt{#1}}\fi

\bibitem{Einav-Levin-2014-Science}
\Name{L. E. \and J. L.} \REVIEW{Science}{346}{2014}{1243089}.

\bibitem{Bouchaud-2008-Nature}
\Name{Bouchaud J.-P.} \REVIEW{Nature}{455}{2008}{1181}.

\bibitem{Schweitzer-Fagiolo-Sornette-VegaRedondo-Vespignani-White-2009-Science}
\Name{Schweitzer F., Fagiolo G., Sornette D., Vega-Redondo F., Vespignani A.
  \and White D.~R.} \REVIEW{Science}{325}{2009}{422}.

\bibitem{Pukthuanthong-Roll-2009-JFE}
\Name{Pukthuanthong K. \and Roll R.} \REVIEW{J. Financ. Econ.}{94}{2009}{214}.

\bibitem{Schiavo-Reyes-Fagiolo-2010-QF}
\Name{Schiavo S., Reyes J. \and Fagiolo G.} \REVIEW{Quant.
  Financ.}{10}{2010}{389-399}.

\bibitem{Battiston-Farmer-Flache-Garlaschelli-Haldane-Heesterbeek-Hommes-Jaeger-May-Scheffer-2016-Science}
\Name{Battiston S., Farmer J.~D., Flache A., Garlaschelli D., Haldane A.~G.,
  Heesterbeek H., Hommes C., Jaeger C., May R. \and Scheffer M.}
  \REVIEW{Science}{351}{2016}{818}.

\bibitem{Boss-Elsinger-Summer-Thurner-2004-QF}
\Name{Boss M., Elsinger H., Summer M. \and Thurner S.} \REVIEW{Quant.
  Financ.}{4}{2004}{677}.

\bibitem{Garlaschelli-Loffredo-2004-PRL}
\Name{Garlaschelli D. \and Loffredo M.~I.} \REVIEW{Phys. Rev.
  Lett.}{93}{2004}{188701}.

\bibitem{Hochberg-Ljungqvist-Lu-2007-JF}
\Name{Hochberg Y.~V., Ljungqvist A. \and Lu Y.} \REVIEW{J.
  Financ.}{62}{2007}{251}.

\bibitem{Kogut-Urso-Wakler-2007-MS}
\Name{Kogut B., Urso P. \and Walker G.} \REVIEW{Manag. Sci.}{53}{2007}{1181}.

\bibitem{Fagiolo-Reyes-Schiavo-2009-PRE}
\Name{Fagiolo G., Reyes J. \and Schiavo S.} \REVIEW{Phys. Rev.
  E}{79}{2009}{36115}.

\bibitem{Kyriakopoulos-Thurner-Puhr-Schmitz-2009-EPJB}
\Name{Kyriakopoulos F., Thurner S., Puhr C. \and Schmitz S.~W.} \REVIEW{Eur.
  Phys. J. B}{71}{2009}{523}.

\bibitem{Tseng-Li-Wang-2010-EPJB}
\Name{Tseng J.-J., Li S.-P. \and Wang S.-C.} \REVIEW{Eur. Phys. J.
  B}{73}{2010}{69}.

\bibitem{Tseng-Lin-Lin-Wang-Li-2010-PA}
\Name{Tseng J.-J., Lin C.-H., Lin C.-T., Wang S.-C. \and Li S.-P.}
  \REVIEW{Physica A}{389}{2010}{1699}.

\bibitem{Jiang-Zhou-2010-PA}
\Name{Jiang Z.-Q. \and Zhou W.-X.} \REVIEW{Physica A}{389}{2010}{4929}.

\bibitem{Wang-Zhou-Guan-2011-PA}
\Name{Wang J.-J., Zhou S.-G. \and Guan J.-H.} \REVIEW{Physica
  A}{390}{2011}{398}.

\bibitem{Tumminello-Lillo-Piilo-Mantegna-2012-NJP}
\Name{Tumminello M., Lillo F., Piilo J. \and Mantegna R.~N.} \REVIEW{New J.
  Phys.}{14}{2012}{013041}.

\bibitem{Sun-Cheng-Shen-Wang-2011-PA}
\Name{Sun X.-Q., Cheng X.-Q., Shen H.-W. \and Wang Z.-Y.} \REVIEW{Physica
  A}{390}{2011}{3427}.

\bibitem{Sun-Shen-Cheng-Wang-2012-PLoS1}
\Name{Sun X.-Q., Shen H.-W., Cheng X.-Q. \and Wang Z.-Y.} \REVIEW{PLoS
  One}{7}{2012}{e45598}.

\bibitem{Jiang-Xie-Xiong-Zhang-Zhang-Zhou-2013-QFL}
\Name{Jiang Z.-Q., Xie W.-J., Xiong X., Zhang W., Zhang Y.-J. \and Zhou W.-X.}
  \REVIEW{Quant. Financ. Lett.}{1}{2013}{1}.

\bibitem{Adamic-Brunetti-Harris-Kirilenko-2012-SSRN}
\Name{Adamic L., Brunetti C., Harris J. \and Kirilenko A.~A.} \Book{{Trading
  networks}} (2010).
\newline\url{http://ssrn.com/abstract=1361184}

\bibitem{Li-Jiang-Xie-Xiong-Zhang-Zhou-2015-PA}
\Name{Li M.-X., Jiang Z.-Q., Xie W.-J., Xiong X., Zhang W. \and Zhou W.-X.}
  \REVIEW{Physica A}{419}{2015}{575}.

\bibitem{Sun-Shen-Cheng-2014-SR}
\Name{Sun X.-Q., Shen H.-W. \and Cheng X.-Q.} \REVIEW{Sci.
  Rep.}{4}{2014}{3711}.

\bibitem{Cohen-Cole-Kirilenko-Patacchini-2011-SSRN}
\Name{Cohen-Cole E., Kirilenko A.~A. \and Patacchini E.} \Book{{How your
  conterparty matters: Using transaction networks to explain returns in CCP
  marketplaces}} http://ssrn.com/abstract=1597738 (2011).

\bibitem{Wood-McInish-Ord-1985-JF}
\Name{Wood R.~A., McInish T.~H. \and Ord J.~K.} \REVIEW{J.
  Financ.}{40}{1985}{723}.

\bibitem{Gallant-Rossi-Tauchen-1992-RFS}
\Name{Gallant A.~R., Rossi P.~E. \and Tauchen G.} \REVIEW{Rev. Financ.
  Stud.}{5}{1992}{199}.

\bibitem{Richardson-Sefcik-Thompson-1986-JFE}
\Name{Richardson G., Sefcik S.~E. \and Thompson R.} \REVIEW{J. Financ.
  Econ.}{17}{1986}{313}.

\bibitem{Chan-Fong-2000-JFE}
\Name{Chan K. \and Fong W.-M.} \REVIEW{J. Financ. Econ.}{57}{2000}{247}.

\bibitem{Lillo-Farmer-Mantegna-2003-Nature}
\Name{Lillo F., Farmer J.~D. \and Mantegna R.} \REVIEW{Nature}{421}{2003}{129}.

\bibitem{Lim-Coggins-2005-QF}
\Name{Lim M. \and Coggins R.} \REVIEW{Quant. Financ.}{5}{2005}{365}.

\bibitem{Zhou-2012-QF}
\Name{Zhou W.-X.} \REVIEW{Quant. Financ.}{12}{2012}{1253}.

\bibitem{Mu-Zhou-Chen-Kertesz-2010-NJP}
\Name{Mu G.-H., Zhou W.-X., Chen W. \and Kert{\'e}sz J.} \REVIEW{New J.
  Phys.}{12}{2010}{075037}.

\bibitem{Zhou-Mu-Kertesz-2012-NJP}
\Name{Zhou W.-X., Mu G.-H. \and Kert{\'e}sz J.} \REVIEW{New J.
  Phys.}{14}{2012}{093025}.

\bibitem{Zhou-2012-NJP}
\Name{Zhou W.-X.} \REVIEW{New J. Phys.}{14}{2012}{023055}.

\bibitem{Beiro-Alvarez-Hamelin-Busch-2008-NJP}
\Name{Beir{\'o} M.~G., Alvarez-Hamelin J.~I. \and Busch J.~R.} \REVIEW{New J.
  Phys.}{10}{2008}{125003}.

\bibitem{Biais-Hillion-Spatt-1995-JF}
\Name{Biais B., Hillion P. \and Spatt C.} \REVIEW{J. Financ.}{50}{1995}{1655}.

\bibitem{Gu-Xiong-Zhang-Chen-Zhang-Zhou-2016-CSF}
\Name{Gu G.-F., Xiong X., Zhang Y.-J., Chen W., Zhang W. \and Zhou W.-X.}
  \REVIEW{Chaos, Solitons \& Fractals}{88}{2016}{48}.

\bibitem{Zhou-Mu-Chen-Sornette-2011-PLoS1}
\Name{Zhou W.-X., Mu G.-H., Chen W. \and Sornette D.} \REVIEW{PLoS
  One}{6}{2011}{e24391}.

\bibitem{Kraus-Stoll-1972-JF}
\Name{Kraus A. \and Stoll H.~R.} \REVIEW{J. Financ.}{27}{1972}{569}.

\bibitem{Dufour-Engle-2000-JF}
\Name{Dufour A. \and Engle R.~F.} \REVIEW{J. Financ.}{55}{2000}{2467}.

\bibitem{Saar-2001-RFS}
\Name{Saar G.} \REVIEW{Rev. Financ. Stud.}{14}{2001}{1153}.

\bibitem{Chiyachantana-Jain-Jiang-Wood-2004-JF}
\Name{Chiyachantana C.~N., Jain P.~K., Jiang C. \and Wood R.~A.} \REVIEW{J.
  Financ.}{59}{2004}{869}.

\end{thebibliography}

\end{document}